\documentclass[journal]{IEEEtran}

\ifCLASSINFOpdf
\else
\fi

\usepackage{graphicx}  
\usepackage{tikz}  
\usepackage{pgfplots} 
\usepackage{epstopdf}  
\usepackage{subfig}    
\usepackage{amsmath}   
\usepackage{boxhandler}  
\usepackage{dsfont}      
\usepackage[flushleft]{threeparttable}   
\let\labelindent\relax 
\usepackage{enumitem}    
\usepackage{booktabs}    
\usepackage{multirow}    
\usepackage{url} 	     
\usepackage{setspace}    
\usepackage{textcomp}    
\usepackage{color}
\usepackage{xcolor}       
\usepackage{booktabs}    
\usepackage{array}       
\usepackage{caption}     
\usepackage[font=footnotesize]{caption}
\usepackage{epsfig}      
\usepackage{mathrsfs}
\usepackage{amsmath}    
\usepackage{amssymb}     
\usepackage{amsxtra}        
\usepackage{verbatim}    
\usepackage{graphics}    
\usepackage{pdflscape}   
\usepackage{enumerate}
\usepackage{enumitem}
\usepackage{setspace}
\usepackage[export]{adjustbox}
\usepackage{setspace}
\usepackage{bm}
\usepackage{lipsum}
\usepackage{algorithm}
\usepackage[noend]{algpseudocode}
\usepackage[noadjust]{cite} 
\usepackage{marginnote}

\usepackage[english]{babel}
\usepackage[intoc, english]{nomencl}
\makenomenclature

\IEEEoverridecommandlockouts                              

\newcommand*{\TitleFont}{%
       \usefont{\encodingdefault}{\rmdefault}{m}{n}%
       \fontsize{27}{29}%
       \selectfont}

\begin{document}  
\bstctlcite{IEEEexample:BSTcontrol} 
       
\title{\TitleFont
Optimal Pacing of a Cyclist in a Time Trial Based on Individualized Models of Fatigue and Recovery
}

\author{Faraz Ashtiani, \textit{Student Member}, \textit{IEEE}, Vijay Sarthy M Sreedhara, Ardalan Vahidi, \textit{Senior Member}, \textit{IEEE}, Randolph Hutchison, Gregory Mocko
%

%
%
\thanks{(\textit{Corresponding author: Faraz Ashtiani})}
\thanks{Faraz Ashtiani ({\tt\footnotesize fashtia@g.clemson.edu}), Vijay Sarthy M Sreedhara ({\tt\footnotesize vsreedh@clemson.edu}), Ardalan Vahidi ({\tt\footnotesize avahidi@clemson.edu}), and Gregory Mocko ({\tt\footnotesize gmocko@clemson.edu}) are with the Department of Mechanical Engineering, Clemson University, Clemson, SC 29634-0921, USA. Randolph Hutchison ({\tt\footnotesize randolph.hutchison@furman.edu}) is with the Department of Health Science, Furman University, Greenville, SC 29613-1000, USA.
}
}

\maketitle
\thispagestyle{plain}
\pagestyle{plain}
\begin{abstract}

This paper formulates optimal pacing of a cyclist on hilly terrain time-trials as a minimum-time optimal control problem. Maximal power of a cyclist serves as a time-varying constraint and depends on fatigue and recovery which are captured via dynamic models proposed early in the paper. Experimental protocols for identifying the individualized parameters of the proposed fatigue and recovery models are detailed and results for six human subjects are shown. In an analytical treatment via necessary conditions of Pontryagin Minimum Principle, we show that the cyclist's optimal power in a time-trial is limited to only four modes of all-out, coasting, pedaling at a critical power, or constant speed (bang-singular-bang). To determine when to switch between these modes, we resort to numerical solution via dynamic programming. One of the subjects is then simulated on four courses including the 2019 Duathlon National Championship in Greenville, SC. The dynamic programming simulation results show 24\% reduction in travel time over experimental results of the self-paced subject who is a competitive amateur cyclist. The paper concludes with description of a pilot lab experiment in which the subject trial time was reduced by 3\% when the near-optimal pace was communicated to her in real-time.


\end{abstract}

\IEEEpeerreviewmaketitle


\nomenclature{\(CP\)}{Critical power}
\nomenclature{\(AWC\)}{Anaerobic work capacity}
\nomenclature{\(c\)}{Pedaling cadence}
\nomenclature{\(P_{adj}\)}{Adjusted recovery power}
\nomenclature{\(P\)}{Cyclist's applied power}
\nomenclature{\(u\)}{Control input variable (cyclist's power)}
\nomenclature{\(w\)}{Remaining anaerobic energy (state variable)}
\nomenclature{\(c_{max}\)}{Maximum pedaling cadence}
\nomenclature{\(c_{max,f}\)}{Maximum pedaling cadence when fully fatigued}
\nomenclature{\(P_{max}\)}{Maximum cyclist power  as a function of $c$ and $w$}
\nomenclature{\(P_{peak}\)}{Peak power as a function of $w$ at optimal cadence}
\nomenclature{\(v\)}{Bicycle's speed (state variable)}
\nomenclature{\(s\)}{Bicycle's position (state variable)}
\nomenclature{\(u_{max}\)}{Maximum constraint on the control input}
\nomenclature{\(C\)}{State and control dependent constraint}
\nomenclature{\(S\)}{State dependent constraint}
\nomenclature{\(\lambda\)}{Vector of the co-state variables}
\nomenclature{\(H\)}{Hamiltonian function}
\nomenclature{\(m\)}{Effective mass of the bicycle}
\nomenclature{\(m_b\)}{Mass of the bicycle and the rider combined}
\nomenclature{\(C_R\)}{Coefficient of rolling resistance}
\nomenclature{\(C_d\)}{Aerodynamic drag coefficient}
\nomenclature{\(\rho\)}{Air density}
\nomenclature{\(A\)}{Frontal area of the bicycle}
\nomenclature{\(a\)}{Recovery model parameter}
\nomenclature{\(b\)}{Recovery model parameter}
\nomenclature{\(\alpha\)}{Maximum power model parameter}
\nomenclature{\(\alpha_{c}\)}{Maximum cadence model parameter}
\nomenclature{\(w_{rec}\)}{Recovered anaerobic energy}
\nomenclature{\(T_{rec}\)}{Recovery duration}
\nomenclature{\(x\)}{Vector of the state variables}
\nomenclature{\(f\)}{Vector of the state equations}
\nomenclature{\(b\)}{Recovery model parameter}
\nomenclature{\(I_w\)}{Rotational inertia of a bicycle's wheel}
\nomenclature{\(R_w\)}{Radius of bicycle  wheel}
\nomenclature{\(g_i\)}{Gear ratio of the $i^{th}$ gear}
\nomenclature{\(L\)}{Integrand in the cost function}
\nomenclature{\(u_{mode}\)}{Optimal mode of control action}
\nomenclature{\(\widehat{CP}\)}{Estimated Critical Power}
\nomenclature{\(\widehat{AWC}\)}{Estimated anaerobic work capacity}
\nomenclature{\(h\)}{The summation of road forces}

\section{INTRODUCTION}
\label{sec_intro}

\IEEEPARstart{O}{ptimization} of human performance by accurately modeling fatigue has challenged athletes, coaches, and scientists. The rise in popularity of wearable sensors in physical activity tracking presents opportunities for modeling and optimizing performance as they alleviate the need for expensive laboratory equipment. Understanding fatigue dynamics can potentially help athletes train in a more efficient way and perform at their peak. Moreover, it can provide useful information to further enhance the performance of an athlete during a physical exercise. Fatigue due to prolonged exercise is defined as a decline in muscle performance which accompanies a sensation of tiredness \cite{abbiss2005models,bigland1984changes}. Therefore, during physical exercise, fatigue prevents athletes from producing the required power. Several studies such as \cite{morton2006critical,bergstrom2014differences} have investigated fatigue in cycling and developed models for it based on the expenditure of anaerobic energy. Only a few studies investigate the recovery dynamics of this anaerobic energy \cite{bogdanis1995recovery}. 

\begin{figure}
\centering
\setlength\fboxsep{0pt}
\setlength\fboxrule{0pt}
\setlength\belowcaptionskip{-10pt}
\fbox{\includegraphics[width=\columnwidth]{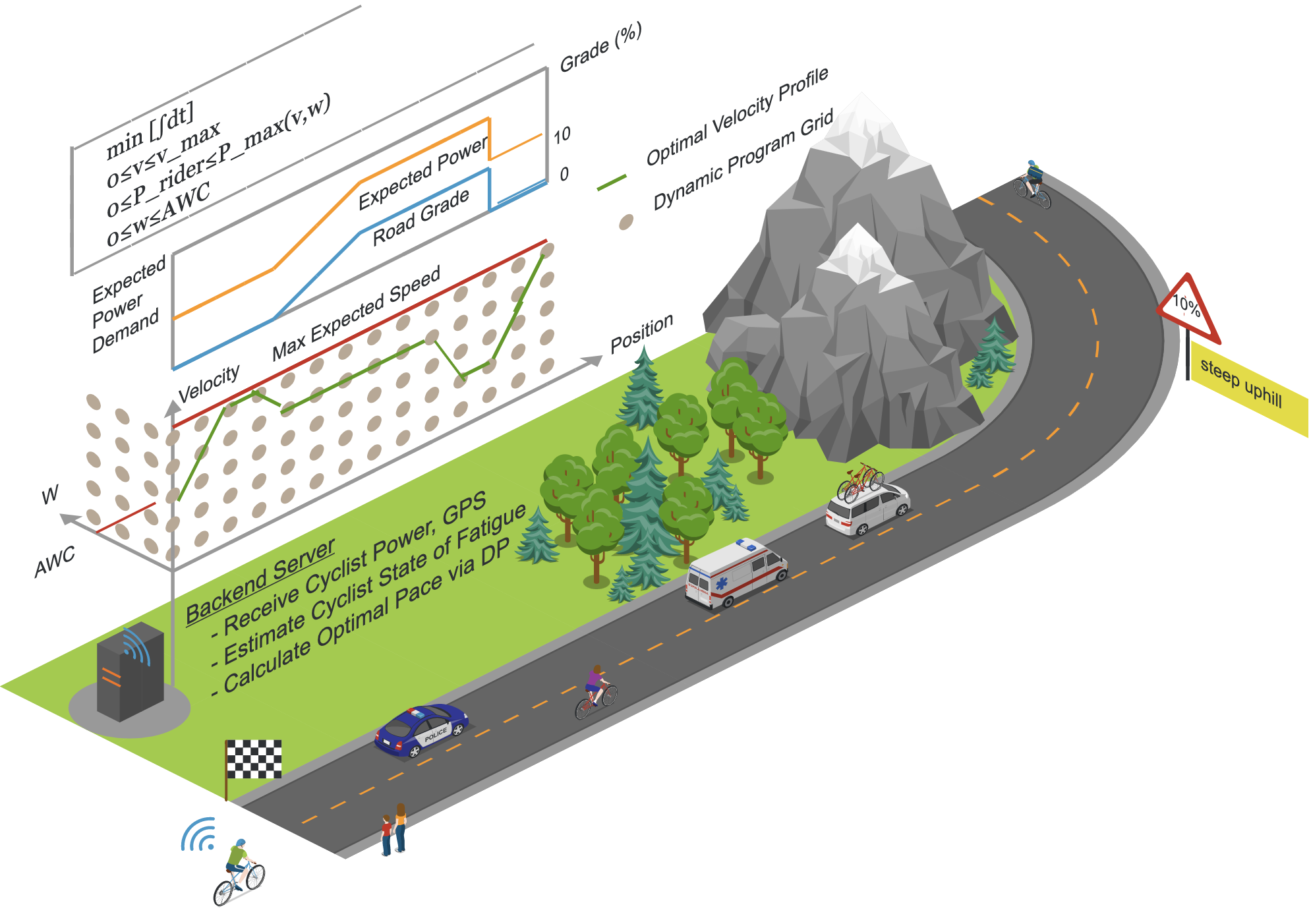}}
\caption{\footnotesize An illustration of optimal pacing of a cyclist using upcoming elevation data, and based on a dynamic model of cyclist fatigue and recovery, and a model of bicycle longitudinal dynamics. The illustration also shows a schematic of a dynamic programming grid on two states of velocity ($v$) and anaerobic energy ($w$) for planning the optimal velocity or power. The optimal power is communicated and displayed to the cyclist in real-time.}
\label{fig_illustration}
\end{figure}    

Figure \ref{fig_illustration} illustrates a cyclist pedaling in a time trail on a mountain course. Even seasoned cyclists may find it challenging to pace themselves in such a course. By over exerting themselves too early or too late, they may not achieve their maximum potential. We hypothesize that cyclists can finish a time trial faster if they plan in advance in consideration of their Anaerobic Work Capacity ($AWC$) and upcoming road elevation. The pacing strategy can be formalized as an optimal control problem which requires dynamic models of fatigue and recovery and maximal power capacity of a cyclist as a function of their fatigue levels. 

There have been a few other studies attempting to formulate a time-trial as an optimal control problem. In \cite{fayazi2013optimal} and \cite{wan2014optimal} authors provide an optimal control formulation that uses lumped muscle models of fatigue and recovery. However, such lumped models are complex and hard to experimentally verify and calibrate. Alternatively, one can consider fatigue as running out of $AWC$ which could be captured with a single dynamic state and recovery is achieved when pedaling below a Critical Power ($CP$). Such a model can be more easily verified and calibrated in laboratory experiments as our recent work in \cite{ashtiani2019experimental} and \cite{vijay2020modeling} shows. In fact, a model based on anaerobic energy expenditure has been used in \cite{de2017individual} to optimize a 5 km cycling time-trial. However, a model for recovery was not considered in \cite{de2017individual} perhaps because the road was assumed to be flat, not necessitating a more sophisticated strategy.  

This paper has the following main contributions:
\begin{itemize}
\item We hypothesize a set of dynamic models for a cyclist's fatigue and recovery during anaerobic exercise and extract the model parameters for six human subjects using 14 hours of experiments per subject via newly proposed test protocols. We believe these models based on anaerobic work capacity are more practical to use in optimal pacing than those based on estimating muscular fatigue presented in \cite{fayazi2013optimal} and \cite{wan2014optimal}. This is mainly because the state of $AWC$ can be estimated rather reliably open-loop and by integrating the cyclist's power output over time  and without the need for invasive (e.g. blood lactate) measurements. 
\item We show that affine dependence of system's dynamics and constraints on control input (pedaling power) limits the optimal power to only four power levels, which is a valuable practical insight into the nature of the optimal pacing strategy when using Pontryagin's Minimum Principle. We resort to numerical solution via Dynamic Programming (DP) to resolve the switching between various modes of optimal strategy considering the varying road grade. 
\item We simulate one of the subjects on four time-trial courses including the 2019 Greenville Duathlon National Championship and show reduced trial time over cyclist's self-paced strategy. Our sensitivity analysis shows robustness to reasonable perturbations in model parameters but also the need for individualized models.
\item A pilot test in which a subject received optimal power suggestions in real-time and improved her trial time over a self-paced baseline trial demonstrates the real-time implementation potential.
\end{itemize}

After the table of nomenclature, Section \ref{sec_lit_review} presents a literature review on muscle fatigue and recovery. Section \ref{sec_modeling} introduces our proposed models for anaerobic energy expenditure and recovery as well as a model for maximum power generation ability of cyclists. In Section \ref{sec_exp_protocol} the experimental protocol and results for six cyclists are presented. The optimal pacing problem is formulated in Section \ref{sec_optimal_formulation}. The problem is investigated analytically in Section \ref{sec_optimal_theory} using optimal control theory, and numerical solutions using dynamic programming are shown in Section \ref{sec_optimal_DP}. Section \ref{sec_optimal_DP_simulation} presents the results followed by conclusions.

\printnomenclature
\section{Literature Review}
\label{sec_lit_review}

\subsection{Fatigue Definition and Mechanism}
\label{Fatigue}

Several studies have investigated fatigue, which has led to its many definitions. For example, in \cite{Vollestad1997measurement} authors define fatigue as a reduction in maximal capacity to generate force or power during exercise. Whereas, fatigue is defined as inability to produce the desired force or power resulting in impaired performance in \cite{edwards1981human} and \cite{enoka1992neurobiology}. The point of occurrence of fatigue is defined as the moment at which a drop from the desired power level is observed, which is also time-to-exhaustion, thus making both terms indistinguishable \cite{fulco1996muscle}. To address this, in \cite{kay2000fluid} and \cite{kay2001evidence} fatigue is defined as a continuous process altering the neuromuscular functional state resulting in exhaustion and exercise termination. In general, there are two sources of fatigue: central fatigue and peripheral fatigue \cite{williams2009human}. Central fatigue is defined as the failure of the Central Nervous System (CNS) in sending necessary commands to muscles to operate \cite{davis1997possible}. This factor is essentially important in high-intensity exercise \cite{kent1999central}. Peripheral fatigue is caused at the muscle level. It can be induced because of the neuromuscular failure in muscles to comprehend and perform the commands coming from CNS, and deficiency of vital substances \cite{kent1999central}.

The above mentioned studies illustrate the difficulty involved in arriving at a global definition of fatigue. The mechanism of fatigue leading to exhaustion are different for cycling, running, swimming, etc \cite{enoka2008muscle}. As we previously presented in \cite{ashtiani2019experimental}, we define muscle fatigue and recovery in cycling as:

\begin{itemize}
\item \textit{Fatigue}: Expending energy from anaerobic metabolic systems by pedaling above a critical power which results in a decrease of maximum power generation ability.
\item \textit{Recovery}: Recuperating energy into anaerobic metabolic systems by pedaling below the critical power which results in an increase of maximum power generation ability. 
\end{itemize}

\subsection{Muscle Fatigue Measurement and Modeling}
\label{Measurement}

Measuring the effect of central fatigue on muscle performance is a challenge since the matter is highly subjective \cite{liepert1996central}. Most research efforts to objectively measure central fatigue are focused on measurement of Maximum Voluntary Contraction (MVC) \cite{moussavi1989nonmetabolic,allman2002neuromuscular,moritani1986intramuscular}. In voluntary contraction of a muscle, the generated force is proportional to the muscle electrical activation \cite{bigland1954motor}. A standard way of measuring muscle activity is via Electromyography (EMG) tests. During the test the amount of electric potential produced in muscles can be measured. Although studies such as \cite{potvin1997validation} and \cite{gerdle2000criterion} have shown the accuracy of EMG tests in measuring maximal and submaximal voluntary contractions, there is evidence of underestimating the muscle activation at high force levels \cite{williams2009human}. Therefore, to point to the goal of the current study, EMG cannot provide accurate data for modeling a cyclist's fatigue and recovery.

In addition to central fatigue there are several sites for peripheral fatigue. To get a better understanding of fatigue at muscle level, we should focus on the muscle metabolic system. Muscle contraction needs a source of energy and the fuel that provides this energy is adenosine triphosphate (ATP). When one phosphate radical detaches from ATP, more than 7300 calories of energy are released to supply the energy needed for muscle contraction \cite{hall2015guyton}. After this detachment, ATP converts to adenosine diphosphate (ADP). When a human muscle is fully rested, the amount of available ATP is sufficient to sustain maximal muscle power for only about 3 seconds, even in trained athletes \cite{hall2015guyton}. Therefore, for any physical activity that lasts more than a few seconds, it is essential that new ATP be formed continuously.

There are three metabolic systems which provide the needed ATP: Aerobic system , Glycogen-lactic acid system and Phosphagen system. Table \ref{tbl:metabolic-systems} compares the three systems, in terms of moles of ATP generation per minute and endurance time at maximal rates of power generation. Utilization of these systems during physical activity is based on the intensity of the activity.

\begin{table}
\begin{center}
\captionsetup{width= \columnwidth}
\caption{\vspace{-0.5mm}\footnotesize Comparison between three metabolic systems that provide ATP for muscle contraction \cite{hall2015guyton}.} \label{tbl:metabolic-systems}
\resizebox{\columnwidth}{!}
{\begin{tabular}{c|c|c}
\hline
	  Metabolic system & Moles of ATP/min & Endurance time \\
\hline\hline
Phosphagen system  &  4 & 8-10 seconds \\
Glycogen-lactic acid system  &  2.5 & 1.3-1.6 minutes \\
Aerobic  &  1 & Unlimited (as long as nutrients last) \\
\hline
\end{tabular}}

\end{center}
\end{table}

Cycling can fall in all categories above depending on the intensity of the exercise. Many people cycle for fun and get around cities. In this case they may only use their aerobic system which provides them with low amount of power which they can hold for a very long time. However, during high intensity cycling such as in a time trial, the human body will use the other two sources besides the aerobic system to provide enough energy for muscle contraction. This is important when hypothesizing mathematical models that describe muscle power generation in cycling. Later on, we discuss a method to define a power limit below which the cyclists use their aerobic metabolic system that allows them to hold their power for long periods of time. However, there is limited energy to pedal above this power limit.

During aerobic exercise, muscles utilize the aerobic metabolic system to produce ATP. Oxygen plays a vital role in formation of ATP molecules. There are two major methods to measure oxygen during a fatiguing exercise. The first one is measuring the volume of oxygen intake in breathing ($\dot{V}_{O_2}$) \cite{cerretelli1979effects}. During a physical exercise, $\dot{V}_{O_2}$ increases to provide the necessary oxygen needed to produce ATP in the muscle as suggested in \cite{barstow1994muscle,saltin1992maximal}, and modeled in \cite{barstow1996influence}. The experimental procedure to measure $\dot{V}_{O_2}$ requires a number of laboratory equipment that cannot be used by a cyclist during everyday outdoor ride. The second method is directly measuring the amount of oxygenated hemoglobins at the local muscle (muscle oxygenation). When muscles are fully fresh, the percentage of oxygenated hemoglobins among the total number of hemoglobins is at its highest. During a fatiguing exercise, this percentage drops. Several studies show that Near Infrared Spectroscopy (NIRS) is a robust method to measure muscle oxygenation \cite{boushel2000near,belardinelli1995changes,van2001performance}. A few companies currently make wearable devices that enable the cyclists to monitor their muscle oxygenation in real-time \cite{moxy,BSX}. We have shown in \cite{ashtiani2019experimental} a real-time measurement of muscle oxygenation during a set of experiments.

Another fatigue indicator is the amount of lactate produced in the muscle. Maximum Lactate at Steady State (\textit{MLSS}) is the maximum maintainable blood lactate concentration without additional buildup through aerobic exercise \cite{beneke2003methodological,billat2003concept}. During an anaerobic exercise, the rate of  lactate production is higher than its dissipation \cite{jacobs1983lactate, friel2012cyclist}. As authors in \cite{hill1927muscular} and \cite{ishii2013effect} suggest, the amount of lactate accumulation can provide useful information about the fatigue level of muscles. The high lactate levels at a muscle represent a lower ability of the muscle to generate force and power \cite{jorfeldt1978lactate, edwards1981human,nummela1992changes}. Traditionally, a common way to measure lactate has been taking blood samples or biopsies during an experiment \cite{aubier1981respiratory,burnley20063}. These invasive methods are not in any way suitable for a real-time measurement and estimation of fatigue. Recently, authors in \cite{mason2018noninvasive} developed a non-invasive method to measure blood lactate using electromagnetic wave sensors. Commercialization of such non-invasive in-situ measurement techniques will enable researchers to develop mathematical models that represent the relationship between blood lactate level and muscle power/force generation capacity.

\subsection{Fatigue in Cycling}
\label{cycling}

In cycling, it is easier to measure power without elaborate laboratory equipment owing to the development of commercial grade power meters. A power meter can be used for training, developing pacing strategies for a time trial, and performance evaluation after an exercise or race \cite{allen2019training}. However, determining exercise intensity using power is not straightforward as a threshold power is needed to classify exercise intensity. As stated in \cite{welch2007affective} and \cite{keir2015exercise}, the Critical Power ($CP$) can potentially be used to determine exercise intensity. It has been shown that $CP$ is close to the power at which MLSS occurs according to \cite{pringle2002maximal,dekerle2003maximal,mattioni2016can}. This means, while pedaling at a power level above $CP$, a cyclist would expend energy from anaerobic energy sources. On the other hand, pedaling below $CP$ helps the cyclist replenish its anaerobic reserve \cite{pringle2002maximal,dekerle2003maximal,mattioni2016can}. 

The critical power concept was introduced by authors in \cite{monod1965work}. They defined $CP$ as the maximum power output that can be maintained indefinitely. In \cite{whipp1982constant} authors showed that there is a limited amount of anaerobic energy for a cyclist to pedal above $CP$. This ``tank'' of energy is called \textit{Anaerobic Work Capacity} ($AWC$). They suggest by pedaling at a certain power level above $CP$, a cyclist can hold that power for a limited amount of time before he or she runs out of anaerobic energy. The relationship between critical power and anaerobic work capacity is often expressed as:

\begin{equation}
\label{eq_basic_AWC}
P = CP + \frac{AWC}{t_{lim}}
\end{equation}

\noindent where $P$ is a constant power level in Watts, and $t_{lim}$ is time-to-exhaustion in seconds. Equation (\ref{eq_basic_AWC}) can be rewritten as $(P-CP)\times t_{lim}=AWC$ which means that the anaerobic energy spent by keeping the power $P$ constant for the duration $t_{lim}$ must equal AWC. The experimental protocol designed to calculate $CP$ and $AWC$ in Equation (\ref{eq_basic_AWC}) requires multiple lab visits for the test subjects according to \cite{bull2000effect,bergstrom2014differences,gaesser1995estimation,hill1993critical,housh1990methodological} . On each test day, the subject is required to pedal at a specified constant power level until exhaustion. Exhaustion happens when a subject pedals below the specified power level for more than 10 seconds straight. This protocol is repeated multiple times at different constant power levels. Equation (\ref{eq_basic_AWC}) shows the relationship between the set constant power level, $CP$, $AWC$, and time to exhaustion during each test.

\begin{figure}
\centering
\setlength\fboxsep{0pt}
\setlength\fboxrule{0pt}
\setlength\belowcaptionskip{0pt}
\fbox{\includegraphics[width=8.5 cm]{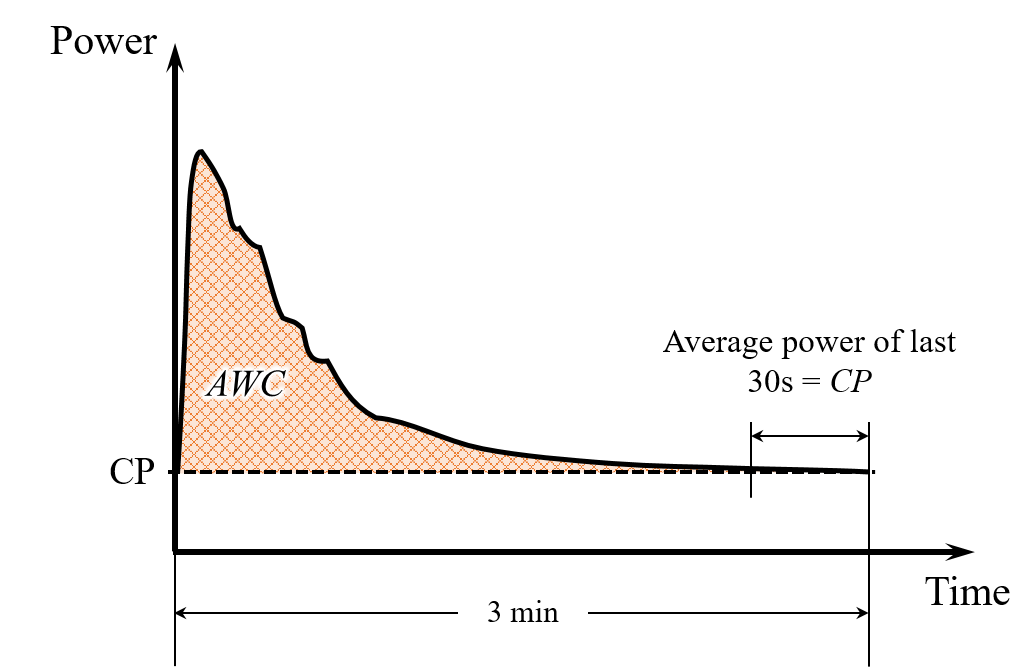}}
\caption{\footnotesize The 3-minute-all-out test protocol. The average power at the last 30 seconds of the tests is considered to be $CP$, and the area between power plot and $CP$ is equivalent to $AWC$.}
\label{fig_3mao}
\end{figure}

To avoid these multiple lab visits, authors in \cite{vanhatalo2007determination} developed a 3-minute-all-out test (3MT) as in Figure \ref{fig_3mao}. In this test, subjects sprint ``all-out'' for the entire 3 minutes. The value of $CP$ is given by the average power output of the last 30 seconds. The value of $AWC$ is the area between the power curve and $CP$. This test has been further validated in \cite{johnson2011reliability} and \cite{vanhatalo2008robustness}. However, there are some recent papers arguing the validity of the 3MT test to estimate $CP$ and $AWC$ \cite{bartram2017predicting, wright2017reliability}. In \cite{bartram2017predicting} authors suggest that while 3MT provides an over estimate of $CP$, it can still be considered as a useful test since it reduces the number of lab visits significantly. For the purpose of this paper, we rely on the 3MT test to determine $CP$ and $AWC$.

\section{Modeling Framework}
\label{sec_modeling}



The focus of this section is i) developing a dynamic model for depletion and recovery of anaerobic work capacity as the cyclist pedals with a power above and below critical power respectively, and ii) determining the maximal power the cyclist can produce as a function of remaining anaerobic work capacity and pedaling cadence.  

\subsection{Dynamic Model of Fatigue and Recovery} As discussed in Section \ref{cycling}, $AWC$ is a finite energy store for pedaling above $CP$. When a cyclist expends his or her $AWC$ entirely, the maximum power that can be produced is $CP$. Let $w$ be the remaining amount of AWC. The rate of change of $w$ with time while expending energy above $CP$ is given by the difference between the rider's power and $CP$, as shown below in Equation (\ref{eq_p_exp_rec}). Recovery of $w$ on the other hand happens when the pedaling power is below $CP$. Recovery rate of $w$ can be calculated similarly to its depletion  \cite{morton2004critical} albeit at a slower rate as shown in \cite{ferguson2010effect} and \cite{bickford2018modeling}. In other words, if the cyclist is pedaling at a constant power level $P$ below $CP$ for the duration $T_{rec}$, the actual amount of recovered energy ($w_{rec}$) will be less than $(CP-P)\times T_{rec}$. Thus, we propose the notion of an \textit{adjusted recovery power} $P_{adj}$ as,

	\begin{equation}
	P_{adj} \ =\ CP-\frac{w_{rec}}{T_{rec}}
	\label{eq_prec}
	\end{equation}

\noindent to be used in a switching model of energy expenditure and recovery as follows,

	\begin{equation}
    \frac{dw}{dt}=\left\{
    \begin{array}{@{} cc @{}}
       -(P-CP) & P \geqslant CP \\
      \\
       -(P_{adj}-CP) & P < CP
    \end{array}\right.
  \label{eq_p_exp_rec}
	\end{equation}

\noindent where $P$ is the pedaling power of the cyclist. We further hypothesize that the recovery power $P_{adj}$ is only a function of pedaling power $P$ and not its duration $T_{rec}$. This assumption stems from the need for a causal energy recovery model. If it is assumed that $w$'s rate of recovery depends on $T_{rec}$, the amount of recovered energy during recovery interval will depend on the recovery duration in the future which does not seem plausible. In our previous study in \cite{ashtiani2019experimental}, we elaborated more on the reason behind making this assumption by using muscle oxygenation data.

Experimental data that is presented later in Section \ref{sec_exp_protocol} shows that $P_{adj}$ can be well approximated as a linear function of pedaling power $P$ with constant $a$ and $b$ that should be identified for each cyclist. Thus in the rest of paper we use,

\begin{equation}
	P_{adj} \ =\ aP + b
	\label{eq_p_adj}
	\end{equation}

\subsection{Model of Maximal Pedaling Power} 
A parameter which is affected by the expenditure of $AWC$ is the cyclist's ability to generate an instantaneous maximum power. In \cite{ashtiani2019experimental} we had shown, using experimental data, the following linear relationship between maximum power $P_{peak}$ and remaining anaerobic energy $w$ during a 3MT test,

\begin{equation}
P_{peak} = \alpha w + CP
\label{eq_old_model}
\end{equation}

\noindent where $\alpha$ is a parameter. We had assumed test subjects were free to choose their pedaling speed (cadence) for maximal power generation. However, several studies \cite{seck1995maximal, sargeant1981maximum, morin2018biomechanics} have shown through a series of short sprints, that cadence affects the maximum power generation ability of cyclists, regardless of their state of fatigue. At any given fatigue state, the maximum power is a parabolic function of cadence as in,

\begin{equation}
P_{max} = 4P_{peak}\left(\frac{c}{c_{max}}\right)\left(1-\frac{c}{c_{max}}\right)
\label{eq_parabolic}
\end{equation}

\noindent where $P_{peak}$ denotes the peak of parabola, $c$ is the cyclist's cadence, and $c_{max}$ is the cyclist's maximum cadence, which is reached when $P_{max}=0$. Other studies investigate also the effect of fatigue on this parabolic relationship \cite{buttelli1996effect, macintosh2011parabolic, hutchison2021effects}. In \cite{hutchison2021effects}, it is shown that the peak of parabolic curves of $P_{max}$ versus pedaling cadence decreases linearly with fatigue state $w$, consistent with our results in \cite{ashtiani2019experimental} and Equation (\ref{eq_old_model}). 

Moreover, maximum cyclist cadence $c_{max}$ also decreases with the state of fatigue $w$ \cite{hutchison2021effects, buttelli1996effect} and this relationship is linear as experimental results in \cite{hutchison2021effects} show. That is,

\begin{equation}
c_{max} = \alpha_{c} w + c_{max,f}
\label{eq_cadence_max}
\end{equation}

\noindent where $\alpha_{c}$ is a model parameter and $c_{max,f}$ is the maximum cadence of the cyclist when $AWC$ is fully depleted. Therefore, we can update Equation (\ref{eq_parabolic}) using Equations (\ref{eq_old_model}) and (\ref{eq_cadence_max}) as,



\begin{equation}
P_{max} = 4(\alpha w+CP)\left(\frac{c}{\alpha_{c} w + c_{max,f}}\right)\left(1-\frac{c}{\alpha_{c} w + c_{max,f}}\right)
\label{eq_parabolic_final}
\end{equation}

\begin{figure}[h!]
\vspace{-0.3in}
\centering
\setlength\fboxsep{0pt}
\setlength\fboxrule{0pt}
\setlength\belowcaptionskip{0pt}
\fbox{\input{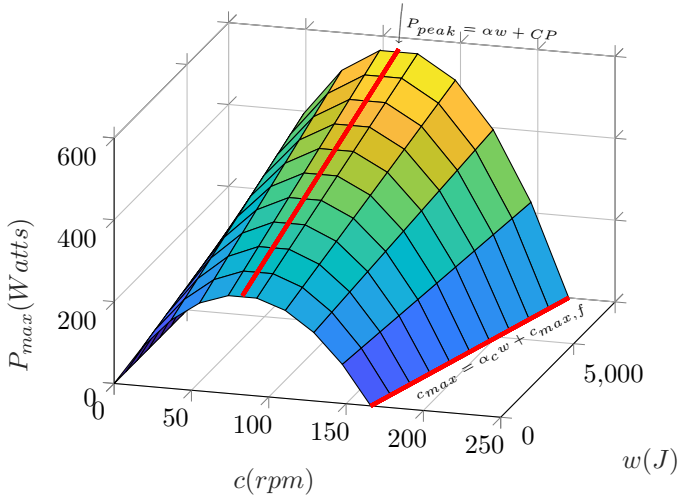}}
\caption{\footnotesize Maximum instantaneous power as a function of the cyclist's cadence and remaining anaerobic energy. The plot is based on data from Subject 14.}
\label{fig_const_cadence}
\end{figure}

\noindent This relationship is visualized in Figure \ref{fig_const_cadence} and was calibrated with data available from one of our human subjects as described in the next section.





\section{Experimental Protocol and Results}
\label{sec_exp_protocol}
The experimental protocol comprised of three tests namely, (i) a ramp test to determine $\dot{V}_{O_2}$ and Gas Exchange Threshold (GET), (ii) a 3-minute all-out test (3MT) to determine $CP$ and $AWC$, and (iii) an interval cycling test to determine the recovery of $AWC$. The ramp test involves incrementally increasing the power at $25 Watts/min$ until the subject is exhausted. From the ramp test, oxygen uptake $\dot{V}_{O_2}$, defined as the volume of oxygen inhaled per minute per kilogram of body weight \cite{cerretelli1979effects}, is determined. Furthermore, during the test, there is an abrupt change in the ratio of volume of $CO_2$ exhaled to the volume of $O_2$ inhaled, which is the GET point at which blood lactate concentration starts to increase. The power at $\dot{V}_{O_2}$ (maximum oxygen uptake) and GET are recorded to aid in designing the 3MT and the interval tests. The $CP$ and $AWC$ from the 3MT are also used to design the interval test for modeling the $AWC$ recovery. The experimental protocol was approved by the Institutional Review Board of Clemson and Furman Universities. Besides power, muscle oxygenation and heart rate were also recorded during the tests.

\begin{figure}[t!]
\centering
\setlength\fboxsep{0pt}
\setlength\fboxrule{0pt}
\setlength\belowcaptionskip{0pt}
\fbox{\includegraphics[width=8.5 cm]{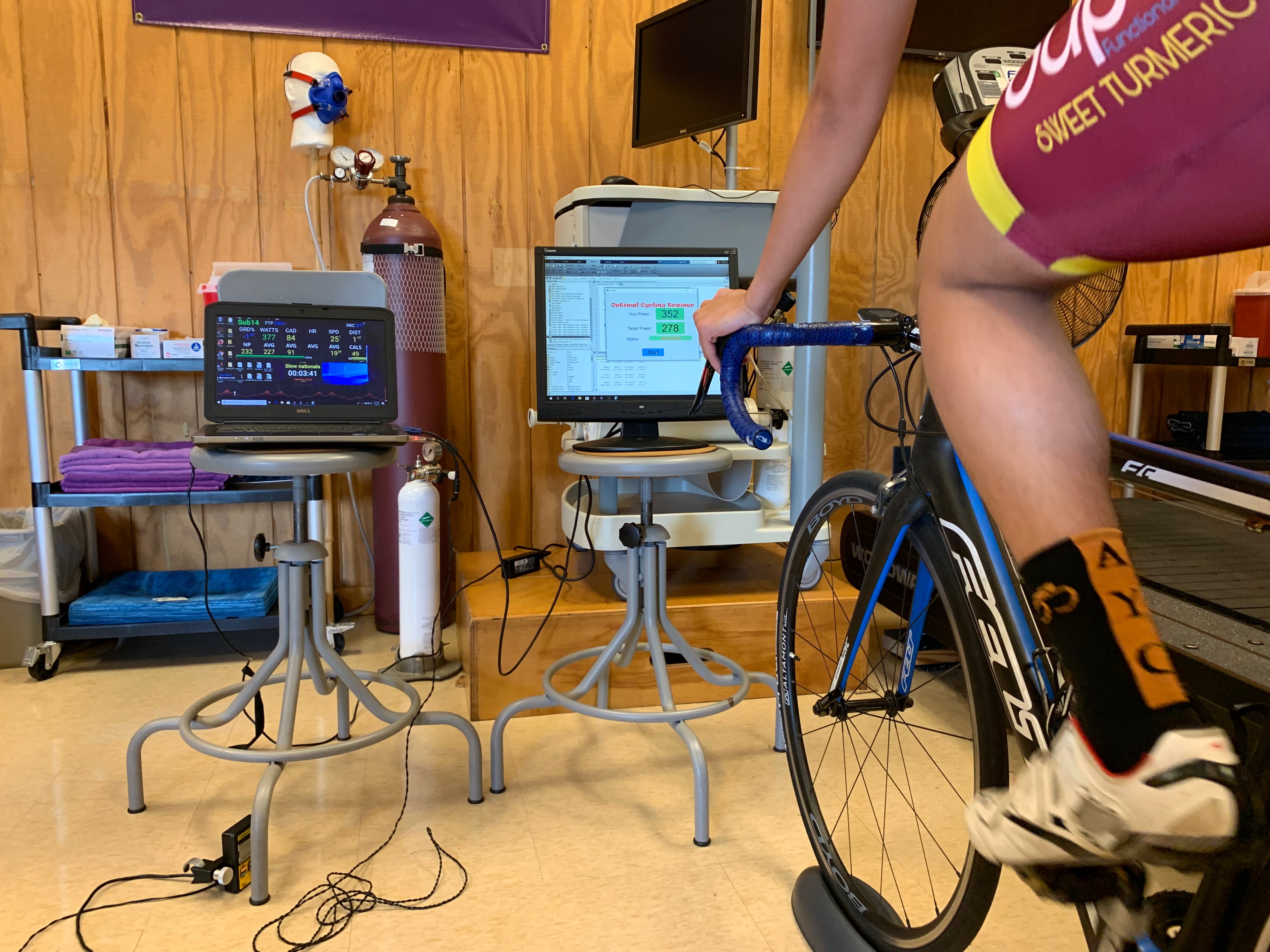}}
\caption{\footnotesize A view of the testing environment. The cyclist in this photo is not one of our subjects.}
\label{fig_experiment_setup}
\end{figure}


We conducted experiments on 17 human subjects. All subjects were cyclists who cycle at least 3 times a week, and were used to high intensity workout sessions. Each subject was scheduled for 14 one-hour-long visits to our laboratory. There was at least 24 hours of resting before each test for the subject to recover to a fully fresh state. Also, the subjects were asked not to consume any caffeine 24 hours prior to each test. The subjects were instructed to remain hydrated during all of the experiments. Because of the complexities of scheduling multiple visits, only 6 of these subjects were able to finish all of the tests. The test protocol was approved by Clemson University's Institutional Review Board (IRB) under protocol numbers IRB2016-169 and IRB2017-222. In order to be compliant with IRB policies we label our subjects by a number. The subjects who successfully finished all the experiments were Subs 6, 9, 11, 12, 14, 16.

All tests were conducted in a laboratory setting on both Clemson and Furman University campuses. Figure \ref{fig_experiment_setup} shows the experimental setup in one of these laboratories. The tests were programmed on a RacerMate CompuTrainer \cite{computrainer} using Perfpro studio software application \cite{perfpro}. There are studies such as \cite{davison2009influence} that suggest the CompuTrainer's power measurement accuracy depends on a variety of parameters such as temperature and calibration procedure. Nevertheless, the CompuTrainer has been shown to be a valid device to estimate $CP$ and $AWC$ using the 3MT test \cite{clark2016validity} and \cite{karsten2015validity}.

\subsection{Estimation of $CP$, $AWC$, and Maximal Power from 3MT}

On the first lab visit, the ramp test was conducted followed by a 3-minute-all-out (3MT) familiarization test. On the next three subsequent visits, a fresh 3MT was conducted from which we averaged estimated values of $CP$ and $AWC$. As shown in Figure \ref{fig_3mao} in each 3MT, subjects sprint ``all-out'' for the entire 3 minutes. The value of $CP$ for each test is given by the average power output of the last 30 seconds as described in \cite{vanhatalo2007determination}. The value of $AWC$ is the calculated area between the power curve and $CP$. We mentioned in Section \ref{sec_lit_review} that there are studies suggesting that the 3MT test may provide an overestimate of $CP$ and $AWC$. Later in the paper, we show via a sensitivity case study that small errors in estimating the values of $CP$ and $AWC$ ($\pm 5\%$) increases the trial time by only $1\%$.

\begin{figure}[h!]
\centering
\setlength\fboxsep{0pt}
\setlength\fboxrule{0pt}
\setlength\belowcaptionskip{-0.2pt}

\fbox{\input{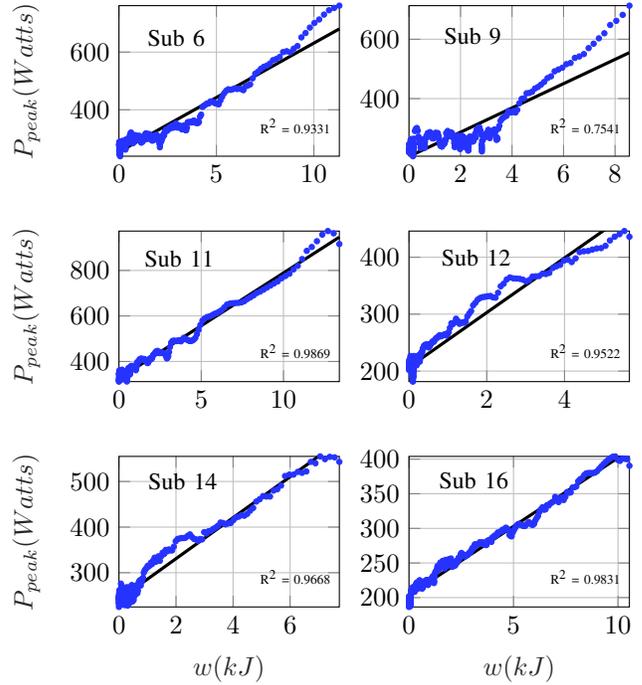}}
\caption{\footnotesize Subjects' peak power versus their remaining anaerobic energy during the 3MT tests.}
\label{fig_peakpower}
\end{figure}

In Equation (\ref{eq_old_model}) we had hypothesized that the peak power $P_{peak}$ during a 3MT can be approximated as a linearly decreasing function of remaining anaerobic energy $w$. Figure \ref{fig_peakpower} shows peak power data as a function of $w$ for all of our six subjects and the $R^2$ as a measure of quality of fit. The close-to-one values of $R^2$ validate the linear relationship for five of the six subjects. The data from subject 9 deviates to some extent from the linear fit; the likely reason being that subject 9 did not perform an all-out effort during the 3MT test. Therefore parameter $\alpha$ in Equation (\ref{eq_old_model}) was estimated by the linear fit to the 3MT peak power data and the result for each subject is reported in Table \ref{tbl:modeling-results}. 

It should be noted that we hypothesized Equation (\ref{eq_parabolic_final}) after we had collected the experimental data. Therefore, we made an assumption to be able to determine the constants $\alpha$, $\alpha_C$ and $C_{max,f}$ in Equation (\ref{eq_parabolic_final}) posthoc. We assume that the braking force applied by the CompuTrainer is set in a way that the subject is always at their ideal cadence and thus at the peak power of the parabolic power-cadence function at each energy level in Figure \ref{fig_const_cadence}. Under this assumption we were able to determine the model constants in Equation (\ref{eq_parabolic_final}) with a single 3MT. Figure \ref{fig_const_cadence} visualizes Equation (\ref{eq_parabolic_final}) using Sub 14's model parameters which will later be used in our optimal control formulation as a constraint.

The summary of modeling results for all of the subjects is presented in Table \ref{tbl:modeling-results}. The reported values for $CP$, $AWC$, and constant parameters of the Equation (\ref{eq_parabolic_final}) are the average of the three 3MT tests each subject performed. An interested reader can find more information about the quality of fit in \cite{vijay2020modeling}.

\begin{table}[t]
\begin{center}
\captionsetup{width= \columnwidth}
\caption{\vspace{0mm}\footnotesize Experimentally determined parameters for the six subjects who successfully finished all of the 3-min-all-out and interval tests.} \label{tbl:modeling-results}
\resizebox{\columnwidth}{!}
{\begin{tabular}{c|c|c|c|c|c|c|c|c|c}
\hline
	  Subject & Sex & m (kg) & $CP$ (Watts) &  $AWC$ (J) & $a$ & $b$ (Watts) & $\alpha$ (1/s) & $\alpha_{c} (rpm/J)$ & $c_{max,f} (rpm)$ \\
\hline\hline
6  & M & 79 & 269 & 12030 & 0.11 & 237.5 & 0.037 & 0.017 & 139\\
9  & M & 63 & 233 & 10100 & 0.09 & 204.5 & 0.036 & 0.014 & 158\\
11  & M & 95 & 335 & 15092 & 0.08 & 300.9 & 0.039 & 0.01 & 163\\
12  & M & 70 & 217 & 5637 & 0.12 & 196.5 & 0.046 & 0.009 & 142\\
14  & F & 74 & 242 & 7841 & 0.08 & 222.5 & 0.044 & 0.008 & 164\\
16  & M & 51 & 206 & 9137 & 0.2 & 167.5 & 0.025 & 0.007 & 154\\
\hline
\end{tabular}}

\end{center}
\end{table}

\subsection{Estimation of Recovery Power from Interval Tests}
\begin{figure*}
\centering
\setlength\fboxsep{0pt}
\setlength\fboxrule{0pt}
\setlength\belowcaptionskip{5pt}
\fbox{\includegraphics[width= 11 cm] {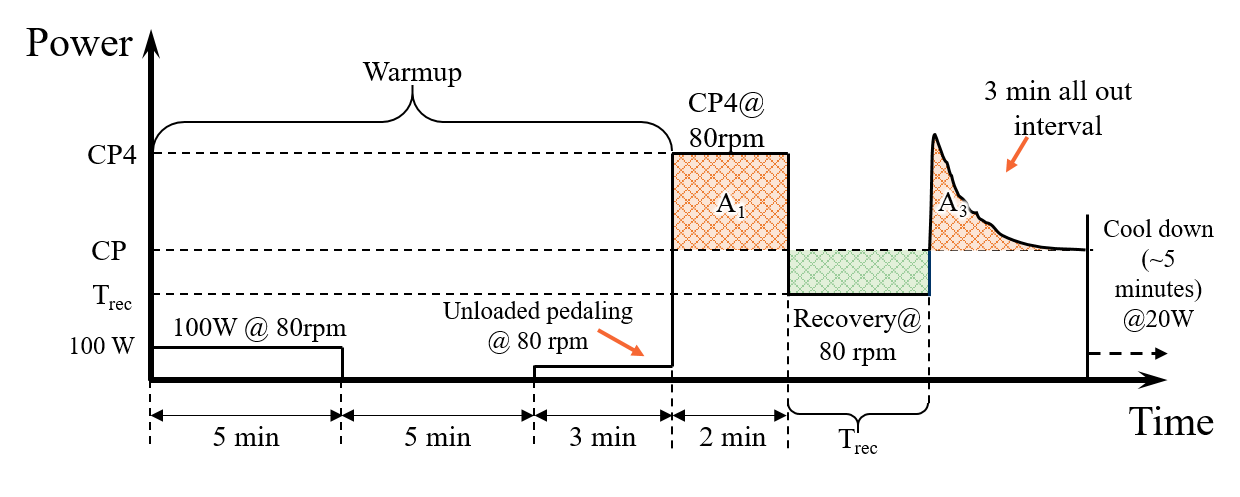}}
\caption{\footnotesize The power interval test protocol. After a warm-up period, the subject pedals at $CP_4$ for 2 minutes. Then the cyclist pedals at three different recovery power levels for different time intervals to recover energy. Following that, the subject performs a 3-min-all-out to burn all the remaining energy from $AWC$.}
\label{fig_interval}
\end{figure*}

The interval test, depicted in Figure \ref{fig_interval}, was developed using the definitions of fatigue and recovery form Section \ref{Fatigue} to derive mathematical models for recovery of $AWC$. The protocol was developed under the following assumptions:

\begin{itemize}
\item The 3MT test provides reliable estimates of $CP$ and $AWC$ \cite{clark2016validity}. 
\item Exercise below $CP$ utilizes the aerobic energy source, and results in recovery of $AWC$ \cite{ferguson2010effect}.
\item The recovery of $AWC$ below $CP$ happens at a slower rate than its expenditure above $CP$ \cite{ferguson2010effect}.
\item The \textit{rate} of recovery depends on the recovery power and not the duration of recovery \cite{ashtiani2019experimental}.
\item The power held during recovery interval is averaged and assumed to be constant.
\end{itemize}

After the 3MT tests and on the fifth visit to the lab, an interval test familiarization was conducted. On the subsequent visits, the interval tests at three different recovery powers (min power on Computrainer 80 Watts, 90\% of power at GET ($P_{GET}$), and half way between $P_{GET}$ and $CP$) and three durations (2 min, 6 min, 15 min) were conducted. The power levels were adopted from \cite{skiba2012modeling} and recovery durations from \cite{ferguson2010effect}. It should be noted that $P_{GET}$ has shown to be always lower than $CP$ \cite{bergstrom2013relationships, constantini2014single}.

In the interval test protocol shown in Figure \ref{fig_interval}, after a warm up of 10 minutes, there is a 2-minute interval at $CP_4$, the power at which all of the subject's $AWC$ will be consumed in 4 minutes ($CP_4 = \frac{AWC}{240}$). The subject expends 50\% of their $AWC$ in the 2-min $CP_4$ interval and then recovers $AWC$ at the above-mentioned recovery powers and durations. Following the recovery interval, the subject then performs a 3-min-all-out test to expend all of their remaining $AWC$. The amount of energy recovered in the recovery interval is then determined by subtracting $AWC$ from the summation of areas above $CP$ through the entire test.

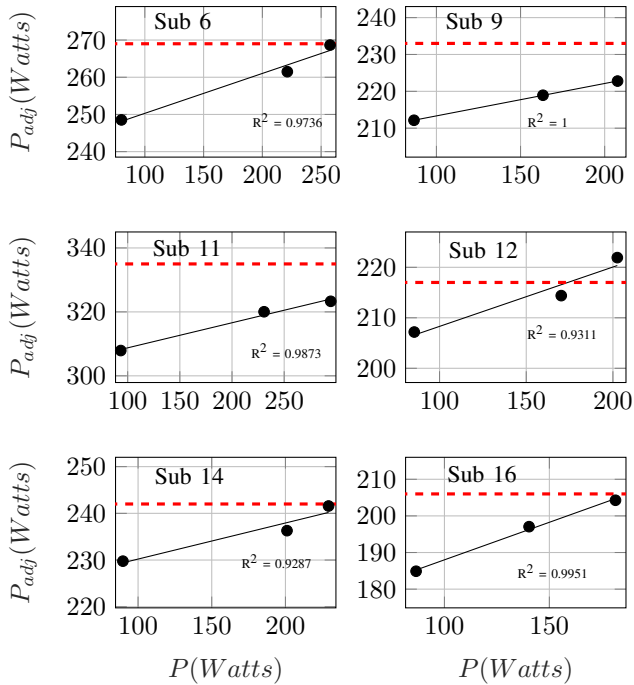
\begin{figure}[t]
\centering
\setlength\fboxsep{0pt}
\setlength\fboxrule{0pt}
\setlength\belowcaptionskip{-0.2pt}
\fbox{
%
%
\definecolor{mycolor1}{rgb}{0.00000,0.44700,0.74100}%
\begin{tikzpicture}

\begin{axis}[%
width=2.93cm,
height=2.003cm,
at={(0cm,5.997cm)},
scale only axis,
xmin=75.06666667,
xmax=262.6,
ymin=238.5290741,
ymax=279,
ylabel style={font=\color{white!15!black}},
ylabel={$P_{adj} (Watts)$},
axis background/.style={fill=white},
xmajorgrids,
ymajorgrids
]
\addplot [color=mycolor1, draw=none, mark size=2.0pt, mark=*, mark options={solid, fill=gray, black}, forget plot]
  table[row sep=crcr]{%
80.06666667	248.5290741\\
221.2333333	261.4898148\\
257.6	268.6433333\\
};
\addplot [color=black, forget plot]
  table[row sep=crcr]{%
80.06666667	248.157955446883\\
257.6	267.202739813198\\
};
\addplot [color=red, dashed, line width=1.2pt, forget plot]
  table[row sep=crcr]{%
56.313333337	269\\
269	269\\
};
\node[right, align=left]
at (axis cs:183.82,248.529) {\tiny $\text{R}^\text{2}\text{ = 0.9736}$};
\node[right, align=left]
at (axis cs:100,275) {\small Sub 6};
\end{axis}

\begin{axis}[%
width=2.93cm,
height=2.003cm,
at={(3.856cm,5.997cm)},
scale only axis,
xmin=81.67,
xmax=212.67,
ymin=202.15,
ymax=243,
axis background/.style={fill=white},
xmajorgrids,
ymajorgrids
]
\addplot [color=mycolor1, draw=none, mark size=2.0pt, mark=*, mark options={solid, fill=gray, black}, forget plot]
  table[row sep=crcr]{%
86.67	212.15\\
163.33	218.94\\
207.67	222.78\\
};
\addplot [color=black, forget plot]
  table[row sep=crcr]{%
86.67	212.163201547077\\
207.67	222.802824325641\\
};
\addplot [color=red, dashed, line width=1.2pt, forget plot]
  table[row sep=crcr]{%
68.57	233\\
225.77	233\\
};
\node[right, align=left]
at (axis cs:148.869,212.15) {\tiny $\text{R}^\text{2}\text{ = 1}$};
\node[right, align=left]
at (axis cs:100,239) {\small Sub 9};
\end{axis}

\begin{axis}[%
width=2.93cm,
height=2.003cm,
at={(0cm,2.998cm)},
scale only axis,
xmin=88.34,
xmax=299.62,
ymin=297.91,
ymax=345,
ylabel style={font=\color{white!15!black}},
ylabel={$P_{adj} (Watts)$},
axis background/.style={fill=white},
xmajorgrids,
ymajorgrids
]
\addplot [color=mycolor1, draw=none, mark size=2.0pt, mark=*, mark options={solid, fill=gray, black}, forget plot]
  table[row sep=crcr]{%
93.34	307.91\\
230.72	320.04\\
294.62	323.31\\
};
\addplot [color=black, forget plot]
  table[row sep=crcr]{%
93.34	308.238081568495\\
294.62	324.015349700781\\
};
\addplot [color=red, dashed, line width=1.2pt, forget plot]
  table[row sep=crcr]{%
67.212	335\\
320.748	335\\
};
\node[right, align=left]
at (axis cs:209.734,307.91) {\tiny $\text{R}^\text{2}\text{ = 0.9873}$};
\node[right, align=left]
at (axis cs:115,340) {\small Sub 11};
\end{axis}

\begin{axis}[%
width=2.93cm,
height=2.003cm,
at={(3.856cm,2.998cm)},
scale only axis,
xmin=80.41,
xmax=207.53,
ymin=197.18,
ymax=227,
axis background/.style={fill=white},
xmajorgrids,
ymajorgrids
]
\addplot [color=mycolor1, draw=none, mark size=2.0pt, mark=*, mark options={solid, fill=gray, black}, forget plot]
  table[row sep=crcr]{%
85.41	207.18\\
170.19	214.39\\
202.53	221.92\\
};
\addplot [color=black, forget plot]
  table[row sep=crcr]{%
85.41	206.582983586102\\
202.53	220.354908733141\\
};
\addplot [color=red, dashed, line width=1.2pt, forget plot]
  table[row sep=crcr]{%
67.698	217\\
217	217\\
};
\node[right, align=left]
at (axis cs:145.271,207.18) {\tiny $\text{R}^\text{2}\text{ = 0.9311}$};
\node[right, align=left]
at (axis cs:100,223.5) {\small Sub 12};
\end{axis}

\begin{axis}[%
width=2.93cm,
height=2.003cm,
at={(0cm,0cm)},
scale only axis,
xmin=84.71,
xmax=234.27,
xlabel style={font=\color{white!15!black}},
xlabel={$P (Watts)$},
ymin=219.76,
ymax=252,
ylabel style={font=\color{white!15!black}},
ylabel={$P_{adj} (Watts)$},
axis background/.style={fill=white},
xmajorgrids,
ymajorgrids
]
\addplot [color=mycolor1, draw=none, mark size=2.0pt, mark=*, mark options={solid, fill=gray, black}, forget plot]
  table[row sep=crcr]{%
89.71	229.76\\
201.07	236.29\\
229.27	241.57\\
};
\addplot [color=black, forget plot]
  table[row sep=crcr]{%
89.71	229.411454024675\\
229.27	240.193614191059\\
};
\addplot [color=red, dashed, line width=1.2pt, forget plot]
  table[row sep=crcr]{%
69.754	242\\
242	242\\
};
\node[right, align=left]
at (axis cs:163.989,229.76) {\tiny $\text{R}^\text{2}\text{ = 0.9287}$};
\node[right, align=left]
at (axis cs:105,248) {\small Sub 14};
\end{axis}

\begin{axis}[%
width=2.93cm,
height=2.003cm,
at={(3.856cm,0cm)},
scale only axis,
xmin=81.49,
xmax=186.51,
xlabel style={font=\color{white!15!black}},
xlabel={$P (Watts)$},
ymin=174.89,
ymax=216,
axis background/.style={fill=white},
xmajorgrids,
ymajorgrids
]
\addplot [color=mycolor1, draw=none, mark size=2.0pt, mark=*, mark options={solid, fill=gray, black}, forget plot]
  table[row sep=crcr]{%
86.49	184.89\\
140.33	197.06\\
181.51	204.27\\
};
\addplot [color=black, forget plot]
  table[row sep=crcr]{%
86.49	185.231493209623\\
181.51	204.7164787374\\
};
\addplot [color=red, dashed, line width=1.2pt, forget plot]
  table[row sep=crcr]{%
70.988	206\\
197.012	206\\
};
\node[right, align=left]
at (axis cs:130.557,184.89) {\tiny $\text{R}^\text{2}\text{ = 0.9951}$};
\node[right, align=left]
at (axis cs:97,211.5) {\small Sub 16};
\end{axis}
\end{tikzpicture}%
}
\caption{\footnotesize Recovery model plots for all six subjects. The gray dots represent the experimental data, the black line shows the model fit, and the red dashed line is $CP$ for each subject.}
\label{fig_sub14rec}
\end{figure}

Since recovery happens at a slower rate than fatigue, we cannot integrate the power curve below $CP$ during the recovery interval (green area in Figure \ref{fig_interval}) to calculate the amount of recovered energy. In Figure \ref{fig_interval}, the sum of orange areas (A$_1$+A$_2$) is the total amount of anaerobic energy expended by the cyclist and an unknown portion of the green area is the recovered energy. The 3-minute-all-out at the end of the protocol, ensures that the anaerobic energy is fully depleted. The total expended energy is more than the subject's $AWC$ because there is some anaerobic energy gained during the recovery interval. Subtract $AWC$ from the total expended anaerobic energy yields the actual amount of recovered energy during the green interval. In other words, the recovered energy is A$_1$+A$_2-AWC$.

In the recovery model we proposed, the goal was to define a substitute power level that could be integrated over time to determine the recovered energy. Therefore, we hypothesized and defined an adjusted power $P_{adj}$ which is computed by dividing the actual recovered energy (A$_1$+A$_2-AWC$) by the recovery duration during a recovery experiment. As explained in Section \ref{sec_modeling}, the recovery rate cannot depend on recovery duration. Therefore, at each power level below $CP$, the adjusted power is calculated for the 3 recovery durations (2 min, 6 min, 15 min) and then averaged. We observed that this adjusted power has an almost linear relationship with the actual applied power for every subject. Figure \ref{fig_sub14rec} shows the three adjusted powers versus the actual pedaling power during recovery intervals for all six subjects which is in agreement with the shown linear fit. The data point close to $CP$ represents the adjusted power for the case where the subject pedals at halfway between GET and $CP$. Since the corresponding adjusted power is very close to $CP$ we expect very little energy recovery at this level. Note that one of the data points for subject 12 is above his $CP$ which could be attributed to the subject deviating from prescribed power during his interval test. Parameters $a$ and $b$ of Equation (\ref{eq_p_adj}) are estimated for each of the six subjects and shown in Table \ref{tbl:modeling-results}.

\section{Optimal Control Formulation}
\label{sec_optimal_formulation}
In this section, pacing of a cyclist is formulated as a constrained optimal control problem. The relevant dynamic states are: i) traveled distance $s$ ii) bicycle velocity  $v$, and iii) the remaining anaerobic energy of the cyclist $w$. The control input is the rider's power at a function of time $u(t)$. Thus the state-space model is of the form:

	\begin{equation}
	\label{eq_states_theory}
	\Dot{x} = f \left(x(t) , u(t) \right) = \begin{bmatrix} f_{1} & f_{2} & f_{3} \end{bmatrix}^T
	\end{equation}
	
\noindent where the state vector is,

\begin{equation}
\label{eq_x}
x(t) = \begin{bmatrix} s(t) & v(t) & w(t) \end{bmatrix}^T \\
\end{equation}

\noindent and $f$ is a nonlinear function, mapping the input and the states to the rate of change of the states. In Equation (\ref{eq_states_theory}), $f_{1}$ is simply velocity,

	\begin{equation}
	\label{eq_state_S}
	\frac{ds}{dt} = v \overset{\Delta}{=} f_{1}
	\end{equation}

\noindent  and $f_{2}$ is obtained using Newton's second law,

	
	\begin{equation}
	\frac{dv(t)}{dt} = \frac{u(t)}{m v(t)} - h(s) - \frac{1}{2m} C_d \rho A v(t)^2  \overset{\Delta}{=} f_2
	\label{eq_state_V}
	\end{equation}

\noindent where

	\begin{equation}
	h(s){=}\frac{m_b}{m} g \left(\sin(\theta)+ C_R \cos(\theta)\right)
	\end{equation}

 \noindent and $m_b$ is the mass of the bicycle and rider, $C_d$ is the aerodynamic drag coefficient, $A$ is the frontal area, $\rho$ is the density of air which is assumed to be constant and independent of the elevation, $\theta$ is the road slope which is positive for uphill and negative for downhill, and $C_R$ is the coefficient of rolling resistance of the road. The parameter $m$ is the effective mass of the bicycle adding the effect of rotational inertia of the wheels to $m_b$  \cite{christie2013lecture},
 
 \begin{equation}
     m =m_b + 2\frac{I_w}{R_w^2}
 \end{equation}

\noindent where $I_w$ is the wheel rotational inertia, and $R_w$ is the radius of each bicycle wheel. The bicycle used in the experiments has a wheel radius of 0.35 $m$ (28 inch wheels) and mass of 1.2 kg, so $I_w$ of each wheel was 0.147 $kg.m^2$. Therefore, the effective mass is 2.3 $kg$ larger than the combined mass of cyclist and the bicycle (2.4\% larger in the case of subject 14).

	
Assuming 100\% efficiency for the bicycle drivetrain, we can equate the propulsion power $u(t)$ to the cyclist's power on the pedals. As a result of using cyclist pedaling power rather than force as input, gear selection is not a factor in our formulation, which otherwise makes the optimization more complex. 

Note that the third state equation for time derivative of cyclist's remaining anaerobic energy is the previously represented Equation (\ref{eq_p_exp_rec}),

\begin{equation}
\label{eq_state_W}
    f_{3} \overset{\Delta}{=} \frac{dw}{dt}=\left\{
    \begin{array}{@{} l c c @{}}
       CP-u & u \geqslant CP & \text{(a)}\\
      \\
       CP-(au + b) & u < CP & \text{(b)}
    \end{array}\right.
\end{equation}

Now we are able to formulate a minimum-time optimal control problem for the pacing strategy in a time-trial. The objective function to be minimized is time,

	\begin{equation}
	\min_{u(t)} \ {J = \int_{t_0}^{t_f} dt}
	\label{eq_t}
	\end{equation}

\noindent subject to,

	\begin{equation}
    \begin{array}{@{} l l @{}}
    \text{state-space model: } & \Dot{x} = f(x(t), u(t))\\
      
      
	  \text{velocity limits: } & 0 \leqslant v(t) \leqslant v_{max}  \\    
      
      \text{remaining energy limits: } & 0 \leqslant w(t) \leqslant AWC \\
      
      \text{rider's power limit: } & 0 \leqslant u(t) \leqslant u_{\,max}(c, w)

    \end{array}
  \label{eq_constraint}
	\end{equation}
 	
\noindent where $u_{max}$ is defined by Equation (\ref{eq_parabolic_final}). In this formulation the final position is specified and fixed but the other two states $v$ and $w$ are let open at the final position. In the simulations of this paper, we assume the maximal speed $v_{max}$ is constant during the trip, but our approach applies if $v_{max}$ varies along the road for example during sharp corners.
\section{Necessary Condition for Optimal Solution}
\label{sec_optimal_theory}

\begin{figure}[h]
\centering
\setlength\fboxsep{0pt}
\setlength\fboxrule{0pt}
\setlength\belowcaptionskip{0pt}
\fbox{\input{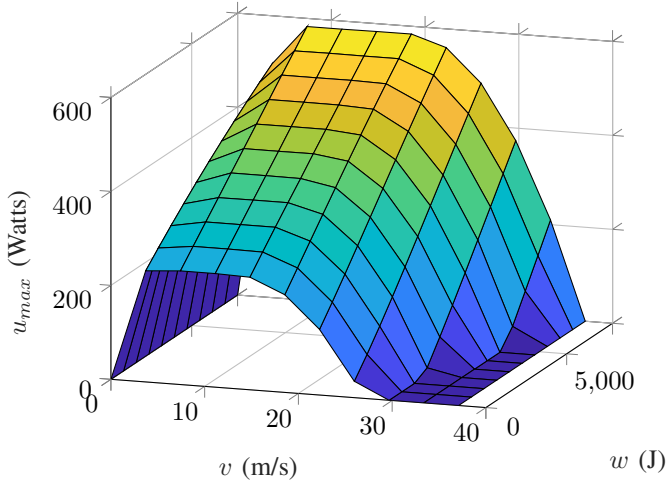}}
\caption{\footnotesize The constraint surface on the control input $u$ as a function of the state variables $v$ and $w$.}
\label{fig_const_final}
\end{figure}

In this section we study the defined optimal control problem using the variational approach. According to Pontryagin's Minimum Principle (PMP), the necessary condition for the optimality of input $u$ is that it minimizes the following Hamiltonian function,

\begin{equation}
\label{eq_hamiltonian}
H(x(t), u(t), \lambda (t)) = L(x(t), u(t)) + \lambda ^{T}(t) \left\lbrace f(x(t),u(t)) \right\rbrace 
\end{equation}

\noindent in which

\begin{equation}
\label{eq_lambda}
\lambda = \begin{bmatrix} \lambda_{1} & \lambda_{2} & \lambda_{3} \end{bmatrix}^T
\end{equation}

\noindent where $\lambda$ is the vector of co-state variables, $L$ is the integrand in the cost function $J$ in (\ref{eq_t}), and $f$ is the vector on the right hand side of the state equations with components represented in Equations (\ref{eq_state_S}), (\ref{eq_state_V}), and (\ref{eq_state_W}). We also need to address the constraint of maximum power generation of the cyclist. The maximum power $u_{max}$ is a function of state variable $w$ and cadence $c$ as represented in (\ref{eq_parabolic_final}). Cadence $c$ needs to be translated to the state variable $v$ for the constraint to be applicable in our formulation. Using the gear ratio of the bicycle in our experimental setup, we can write the bicycle's velocity $v$ at each gear as,

\begin{equation}
v(t) = g_i R_w c
\label{eq_gear_ratio}
\end{equation}

\noindent where $g_i$ is the gear ratio of the $i$\textsuperscript{th} gear, and $R_w$ is the rear wheel (tire thickness included) radius. Our lab's bicycle has 21 gear combinations which means we will have 21 plots in the form of Figure \ref{fig_const_cadence} with different maximum velocities. We take the maximum surface of all of these 21 plots and create the constraint surface as a function of the two states of $v$ and $w$ as shown in Figure \ref{fig_const_final}. The underlying assumption in forming this maximal constraint is that given a specific feasible power to hold, the cyclists can find the best gear that enables them to do so. This is a reasonable assumption because seasoned cyclists are experienced in finding appropriate gears.

Therefore, we observe an inequality constraint as a function of the control input $u$, the state $v$ and the state $w$ variables. We can rewrite the constraint in a standard form as,

\begin{equation}
\label{eq_constraint_u}
C(w,u) = u(t) - u_{max}(v(t), w(t)) \leqslant 0 
\end{equation}

\noindent where $u_{max}(v, w)$ is defined by the surface in Figure \ref{fig_const_final}. 

In the presence of such an inequality constraint on control and state variables, the Hamiltonian equation needs to be augmented as follows \cite{bryson2018applied},

\begin{equation}
\label{eq_hamiltonian_init}
H = L + \lambda^T f + \mu C
\end{equation}

\noindent where 



\begin{equation}
     \mu
   \left\{
    \begin{array}{@{} l c @{}}
     \geqslant 0 & C = 0\\
    \ \ \\
     = 0 & C < 0
       
    \end{array}\right.
\end{equation}

Additionally, the upper and lower limits on $v$ and $w$ presented in Equation (\ref{eq_constraint}) should be taken into account as,

\begin{equation}
S(v,w) = \begin{bmatrix} v(t)-v_{max} \\ -v(t) \\ w(t)-AWC \\ -w(t) \end{bmatrix} \leqslant \vec{0}
\label{eq_constraint_v_w}
\end{equation}

According to \cite{bryson2018applied}, when constraints are not functions of the control input $u$, we take successive time derivatives of $S(v,w)$ and replace $\Dot{v}$ and $\Dot{w}$ with $f_2$ and $f_3$, respectively, until we obtain an expression that is explicitly dependent on $u$. In this case, the first derivative of $S(v,w)$ with respect to both $v$ and $w$ includes $u$. Then, we can treat $S^{(1)}(v,w)$ similar to $C(v,w)$ and augment the Hamiltonian as follows,

\begin{equation}
\label{eq_hamiltonian_revised}
H = L + \lambda^T f + \mu C + \eta^T S^{(1)}
\end{equation}

\noindent where $\eta = \begin{bmatrix} \eta_1 & \eta_2 & \eta_3 & \eta_4 \end{bmatrix}^T$ must obey the following conditions,

\begin{equation}
     \eta_i
   \left\{
    \begin{array}{@{} l c @{}}
     \geqslant 0 & S^{(1)}_i = 0\\
    \ \ \\
     = 0 & S^{(1)}_i < 0
    \end{array}\right.\ \ \ \ \ \text{for}\ \ i = 1,2,3,4
\end{equation}

It should be noted that the terms $\mu C$ and $\eta^T S^{(1)}$ will always be zero in the Hamiltonian when constraints are met. The dynamics of the co-states follow,

\begin{equation}
\label{eq_euler_lagrange}
\dot{\lambda}^T = -H_x \equiv -L_x - \lambda^T f_x - \mu C_x - \eta^T S^{(1)}_x
\end{equation}

\noindent where the subscript $x$ denotes partial derivative with respect to the corresponding state variables. Because dynamics of $w$ in Equation (\ref{eq_state_W}) switches between fatigue and recovery conditions, we end up with a switching Hamiltonian function for fatigue and recovery modes. 

A necessary condition for optimality is that the control input $u$, minimizes the Hamiltonian. One can set the partial derivative of $H$ with respect to $u$ equal to zero. However, in this case because $H$ is affine in $u$, the derivative with respect to $u$ ($H_u$) does not depend on $u$.

\begin{equation}
\label{eq_hamiltonian_u}
H_u = 
\left\{
    \begin{array}{@{} l c @{}}
     \frac{\lambda_2}{mv} - \lambda_3 + \frac{1}{mv}(\eta_1 - \eta_2) + (\eta_4 - \eta_3) + \mu & u \geqslant CP\\
	\ \ \\	    
     \frac{\lambda_2}{mv} - a\lambda_3 + \frac{1}{mv}(\eta_1 - \eta_2) + a(\eta_4 - \eta_3) & u < CP
       
    \end{array}\right.
\end{equation}

Note that the $\mu$ term only shows up when $u \geqslant CP$ because when $u < CP$ the constraint $C$ is not active. Since Equation (\ref{eq_hamiltonian_u}) does not depend on $u$, the optimal solution will be of the bang-singular-bang form; that is the Hamiltonian is minimized at extreme values of $u$ with the exception of potential singular arcs in between.


%
%
%
%
%
%
%


When the Hamiltonian is affine in $u$, the sign of the $H_u$ indicates the optimal input value. As is shown in Equation (\ref{eq_hamiltonian_u}), the sign of $H_u$ depends on $\mu$ and $\eta$. When $\mu$ has a non-zero (positive) value, the constraint $C$ in Equation (\ref{eq_constraint}) is active which means the optimal value for $u$ is its maximum ($u_{max}$) regardless of the sign of the slope. When either $\eta_1$ or $\eta_2$ has a non-zero value, $S^{(1)}_1$ or $S^{(1)}_2$ is active. In that case, it can be shown that acceleration should be zero which means the optimal value for $u$ is the power at which velocity is constant ($u_{\Dot{v}=0}$). When either $\eta_3$ or $\eta_4$ has a non-zero value, $S^{(1)}_3$ or $S^{(1)}_4$ is active. In this case, it can be shown that $\Dot{w}$ is zero which means the optimal input $u$ will be at $CP$. The only cases left are when $\mu$ and $\eta$ are all zero and we can cross them off from Equation (\ref{eq_hamiltonian_u}) and rewrite,

\begin{equation}
H_u = 
\left\{
    \begin{array}{@{} l c @{}}
     \frac{\lambda_2}{mv} - \lambda_3  & u \geqslant CP\\
	\ \ \\	    
     \frac{\lambda_2}{mv} - a\lambda_3 & u < CP
       
    \end{array}\right.
\end{equation}
 
The system of equations that should be solved are the Equations (\ref{eq_states_theory}) and (\ref{eq_euler_lagrange}), resulting in 6 equations combined. The expanded version of the Equation (\ref{eq_euler_lagrange}) will be,

\begin{equation}
\label{eq_extended_euler}
\left\{
\begin{array}{l}
\dot{\lambda_1} = \lambda_2 g\frac{m_b}{m}\frac{d\theta(s)}{ds}\left(\cos(\theta(s)) - C_R\sin(\theta(s))\right) \\
 \\
\dot{\lambda_2} = -\lambda_1 - \frac{\lambda_2}{m_t} \left(\frac{u}{v^2} - C_d \rho A v \right) \\
 \\
\dot{\lambda_3} = 0\\

\end{array} \right.
\end{equation}



Now $H_u$ can be compared in both fatigue and recovery modes. If $H_u$ is positive, the minimum value of $u$ minimizes the function. On the other hand, if $H_u$ is negative, the maximum value of $u$ minimizes the function. We can consider four cases,
\\

\noindent \textbf{Case I.} $\frac{\lambda_2}{mv} - \lambda_3 < 0$ AND $\frac{\lambda_2}{mv} - a\lambda_3 < 0$

The slope in both fatigue and recovery modes are negative which means maximum value of $u$ in each case minimizes the Hamiltonian. In the fatigue mode the maximum input value is $u_{max}$, and in the recovery mode the maximum input value is $CP$. The optimal between the two is the control that yields the smallest Hamiltonian,

\begin{equation}
H^* =\min \left\{ H_{\text{fatigue}}(u_{max}), \ \ H_{\text{recovery}}(CP) \right\}
\end{equation}

\noindent where the subscripts of $H$ differentiates between fatigue and recovery modes because the equations of the two modes are slightly different. If we substitute the aforementioned values of $u$ we cannot decisively say which Hamiltonian is smaller. We can write the optimal input as,

\begin{equation}
\label{eq_case1}
u^* = 
\left\{
    \begin{array}{@{} l c @{}}
     u_{max} & H_{\text{fatigue}}(u_{max}) < H_{\text{recovery}}(CP)\\
	\ \ \\	    
     CP & H_{\text{fatigue}}(u_{max}) > H_{\text{recovery}}(CP)
       
    \end{array}\right.
\end{equation}

Note that in the event $H_{\text{fatigue}}(u_{max}) = H_{\text{recovery}}(CP)$, both $u=CP$ and $u=u_{max}$ are optimal. 
 \\
 
\noindent\textbf{Case II.} $\frac{\lambda_2}{mv} - \lambda_3 > 0$ AND $\frac{\lambda_2}{mv} - a\lambda_3 > 0$

In this case, the input should take its minimum value in both cases, which will be $CP$ and 0 for fatigue and recovery modes, respectively. The minimum Hamiltonian can be found from,

\begin{equation}
H^* =\min \left\{ H_{\text{fatigue}}(CP), \ \ H_{\text{recovery}}(0) \right\}
\end{equation}

\noindent The optimal input value in this case will be:

\begin{equation}
\label{eq_case2}
u^* = 
\left\{
    \begin{array}{@{} l c @{}}
     CP & H_{\text{fatigue}}(CP) < H_{\text{recovery}}(0)\\
	\ \ \\	    
     0 & H_{\text{fatigue}}(CP) > H_{\text{recovery}}(0)
       
    \end{array}\right.
\end{equation}

If $H_{\text{fatigue}}(CP) = H_{\text{recovery}}(0)$, both $u=0$ and $u=CP$ are optimal.
\\

\noindent\textbf{Case III.} $\frac{\lambda_2}{mv} - \lambda_3 > 0$ AND $\frac{\lambda_2}{mv} - a\lambda_3 < 0$

In this case, the input takes its minimum value in fatigue mode, and its maximum value in recovery mode. In both of these scenarios the input is $CP$,

\begin{equation}
u^* = CP
\end{equation}

\noindent \textbf{Case IV.} $\frac{\lambda_2}{mv} - \lambda_3 < 0$ AND $\frac{\lambda_2}{mv} - a\lambda_3 > 0$

In this case, the input takes its maximum value in fatigue mode, and its minimum value in recovery mode, which will be $u_{max}$ and $0$, respectively. The minimum Hamiltonian can be found from,
\begin{equation}
H^* =\min \left\{ H_{\text{fatigue}}(u_{max}), \ \ H_{\text{recovery}}(0) \right\}
\end{equation}
Th optimal input value in this case will be:

\begin{equation}
\label{eq_case3}
u^* = 
\left\{
    \begin{array}{@{} l c @{}}
     u_{max} & H_{\text{fatigue}}(u_{max}) < H_{\text{recovery}}(0)\\
	\ \ \\	    
     0 & H_{\text{fatigue}}(u_{max}) > H_{\text{recovery}}(0)
         \end{array}\right.
\end{equation}

And when $H_{\text{fatigue}}(u_{max}) = H_{\text{recovery}}(0)$, both $u=0$ and $u=u_{max}$ are optimal.

So far the optimal control input can take values of $0$, $CP$, and $u_{max}$. There is yet another case, a possible singularity condition, that will be investigated next.
\\

\noindent \textbf{Singular Arc.} $\frac{\lambda_2}{mv} - \lambda_3 = 0$ OR $\frac{\lambda_2}{mv} - a\lambda_3 = 0$

Here we present the analysis for the case where the Hamiltonian's slope in the recovery or fatigue mode is zero and thus we may have a singular interval which could be optimal. It can be shown that the final results for both modes are the same. Therefore, we only present the analysis for the recovery mode. During the possible singular condition in recovery mode we have,

\begin{equation}
\label{eq_sing_cond}
\frac{\lambda_2}{mv} - a\lambda_3 = 0
\end{equation}

This equality needs to hold for an interval of time for a singular interval to exist. Therefore, its time derivative during that interval should also be zero. Setting the time derivative equal to zero and observing from Equation (\ref{eq_extended_euler}) that $\Dot{\lambda}_3=0$ we get,

\begin{equation}
\frac{\Dot{\lambda}_2}{v} - \frac{\lambda_2}{v^2}\Dot{v} = 0
\end{equation}

We can substitute $\Dot{\lambda}_2$ and $\Dot{v}$ from Equations (\ref{eq_extended_euler}) and (\ref{eq_state_V}) respectively, to obtain,

\begin{equation}
\begin{split}
& \frac{1}{v} \bigg[ -\lambda_1 + \lambda_2 \frac{u}{mv^2} + \frac{\lambda_2}{m}(C_d \rho A)v \\
& - \frac{\lambda_2}{v} \left( \frac{u}{mv} - h(s) - \frac{1}{m}(C_d \rho A)v^2 \right) \bigg]=0
\end{split}
\end{equation} 

Simplifying the equation above by using (\ref{eq_sing_cond}) yields,

\begin{equation}
\lambda_1 + am\lambda_3 h(s) +\frac{3}{2}(C_d \rho A)a\lambda_3 v^2 = 0
\label{eq_firs_derivative}
\end{equation}

Since input $u$ does not appear in Equation (\ref{eq_firs_derivative}), it is necessary to take the time-derivative again which yields,

\begin{equation}
\begin{split}
& -\Dot{\lambda}_1 + am \Dot{\lambda}_3 h(s) + am\lambda_3 \frac{dh}{ds}v \\
& + \frac{3}{2}a\Dot{\lambda}_3(C_d \rho A)v^2 + 3a\lambda_3(C_d \rho A)v\Dot{v} = 0
\end{split}
\end{equation}

Simplifying by using Equation (\ref{eq_extended_euler}) yields,

\begin{equation}
3a\lambda_3(C_d \rho A)v\Dot{v} = 0
\label{eq_sing_final}
\end{equation}

In Equation (\ref{eq_sing_final}) only $\Dot{v}$ can be zero. Therefore, during a potential singular interval the velocity must be a constant. The corresponding input power is obtained using Equation (\ref{eq_state_V}),

\begin{equation}
u^* = u_{\Dot{v}=0} =mv \left( h(s)+ \frac{1}{2m}(C_d \rho A)v^2 \right)
\end{equation}

\noindent which varies with the road grade. 

Considering all of the cases discussed above, the optimal power trajectory can only take values from the vector below,

\begin{equation}
u^* = \begin{bmatrix} u_{max} \\ u_{\dot{v}=0} \\ CP \\ 0 \end{bmatrix}  
\label{eq_bang_bang}
\end{equation}

Equation (\ref{eq_bang_bang}) provides the \textit{necessary} conditions for optimality of $u$ since it was based on a PMP analysis. It therefore suggests that the global minimizer at each time $u^*(t)$ must be one of these modes. This indicates that the optimal pedaling strategy may include one or more of the four modes of pedaling at maximal power, riding at $CP$, resting, or riding at a constant speed. While this insight is useful, it is difficult to analytically determine when the switching between these modes happens. Note that the three state equations along with the three co-state equations form a two point boundary value problem that is generally difficult to solve analytically. The switching dynamics of the problem at hand creates additional challenges. Therefore, next we resort to numerical solution of the optimal control problem via dynamic programming. The above PMP analysis proves valuable in limiting the input space in DP process to only the modes described by Equation (\ref{eq_bang_bang}) which significantly reduces the computational burden of DP.

\section{Numerical Solution of the Problem}
\label{sec_optimal_DP}
In the DP implementation, we use distance as the independent variable instead of time because the final distance is given and road grade is also known as a function of distance. The distance is discretized at equal sampling intervals. With a zero-order hold on input in between sampling intervals, the following discretized state-space equations are obtained,

\begin{equation}
\label{eq_discrete_S}
t_{i+1} = t_i + \frac{\Delta s}{v_i}
\end{equation}

\begin{equation}
	\begin{split}
	&v_{i+1} \ = \ v_i + \frac{\Delta s}{v_i} \bigg(\frac{u_i}{m_t v_i} - g (sin(\theta_i) + \mu cos(\theta_i)) \\
	& - \frac{0.5 C_d \rho A}{m_t} v_i^2 \bigg)
	\label{eq_newton_dis}
	\end{split}
\end{equation}


\begin{equation}
    \left\{
    \begin{array}{@{} l c @{}}
      w_{i+1} = w_i +\frac{\Delta s}{v_i} (CP-u_i)& u_i \geqslant CP\\
      \\
      w_{i+1} = w_i + \frac{\Delta s}{v_i} (CP-au_i-b)& u_i < CP
    \end{array}\right.
  \label{eq_w_dis}
	\end{equation}

Furthermore, the states $v$ and $w$ are quantized in a fine grid between their minimum and maximum (see a schematic illustration of a partial DP grid in Figure \ref{fig_illustration}). The control variable $u$ is only needed to be quantized at the four possible optimal modes determined in Equation (\ref{eq_bang_bang}), which significantly reduces the required memory needs and computation time. 

The cost function in Equation (\ref{eq_t}) is rewritten with position as the independent variable and discretized as follows\footnote{When implementing this objective function in DP, we sometimes observed chattering in the power trajectory. This chattering can be the optimal solution or due to the coarse quantization of control input to only four (optimal) modes. To achieve a practical power trajectory for a cyclist, we added a regularization term to the cost that penalizes the change in control input from stage to stage multiplied by a small penalty weight. The weight was carefully tuned to minimize the impact of the regularization term on the value of the objective function.}, 

\begin{equation}
\label{eq_cost_discrete}
J_N = \sum_{i=0}^{i=N} \frac{\Delta s_i}{v_i}
\end{equation}


\begin{figure*}[h]
\centering
\setlength\fboxsep{0pt}
\setlength\fboxrule{0pt}
\setlength\belowcaptionskip{0pt}
\fbox{\input{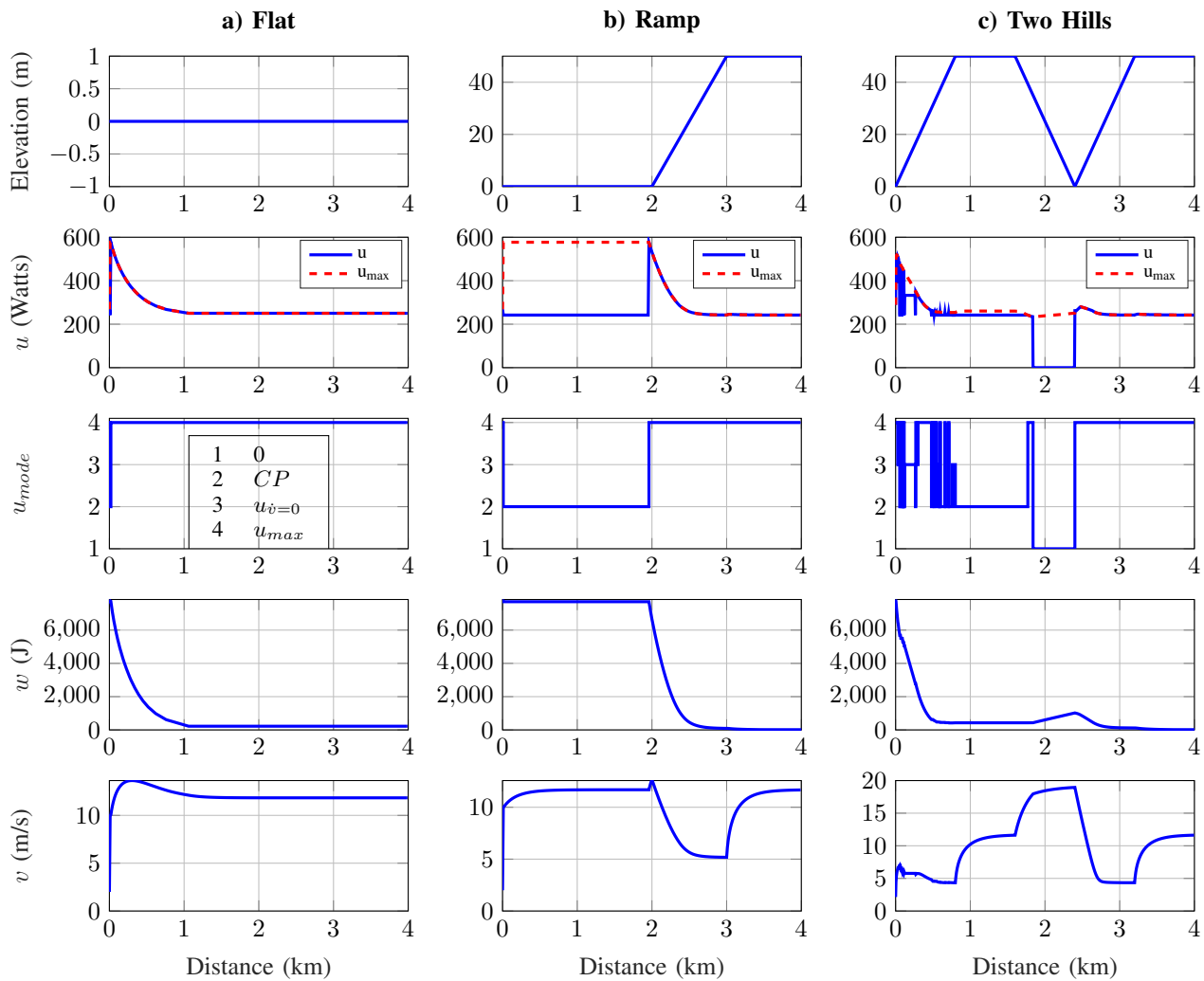}}
\caption{\footnotesize DP simulation results over three elevation profiles. Sub14 model was used in this set of simulations.}
\label{fig_sim_result_test}
\end{figure*}

According to the Bellman's principle of optimality \cite{bellman2015applied}, when a system is on its optimal path from an initial state to a final state, regardless of any past decision or state, it should follow an optimal policy for the remainder of the route. Therefore, in dynamic programming, to find the optimal state trajectory, one can begin from the final state and move backward and calculate the optimal cost-to-go from any state to the final step. 
The optimal costs, as well as optimal control inputs, from all of the possible ($v, w$) nodes at $s_{i+1}$ to the final state at $s_N$ are stored as $J^*_{i+1,N}$. Then, cost-to-go from every node at $s_i$ to all of the nodes at $s_{i+1}$ is calculated and the optimal value among them and the corresponding optimal input is stored for the specified node at $s_i$. This process is repeated backwards to the beginning of the route. As a result a lookup state feedback map is created offline from which the optimal control action can be retrieved for any state at any position stage in real-time. Subsequently, in a forward DP sweep, the optimal action at each discretized state node will be known. Any deviation from the originally planned optimal path can be handled by looking up the optimal input as a function of estimated state from the stored DP map. Let's denote the optimal cost-to-go from specific velocity and energy states at step $s_{i+1}$ to $s_N$ by $J^*_{i+1,N}$. Then, the optimal cost from step $s_i$ to $s_N$ will be,

\begin{equation}
\label{eq_cost_DP_theory}
J^*_{i,N} = \min_{u_i} \ \  [J_{i,i+1} + J^*_{i+1,N}(x)]
\end{equation}

\noindent where

\begin{equation}
\label{eq_cost_DP_theory2}
J^*_{i+1,N} = \min_{{u_{i+1},u_{i+2}, ..., u_{N-1}}} \ \ [J_{i+1,N}]
\end{equation}

\noindent and the optimal control $u^*(i)$ is the minimizing argument. 

In our implementation, the distance is discretized at intervals of $\Delta s = 10$ m. At every distance stage, velocity $v$ is quantized to 300 nodes spaced uniformly in the interval [1, 20] $m/s$. The choice of minimum velocity at $1 m/s$ is because i) speeds closer to zero cause numerical issues as $v$ appears in the denominator in the discretized equations of motion, and ii) during a time-trial the athlete is unlikely to pedal at lower speeds. Anaerobic energy $w$ is quantized to 600 nodes spread uniformly between $0$ and the Subject 14's $AWC$ ($7841 J$). As mentioned earlier, the input power $u$ is also quantized to 4 values as indicated by Equation (\ref{eq_bang_bang}). 

When at the distance $s_i$ from the initial position, $v_i$ and $w_i$ states move to $v_{i+1}$ and $w_{i+1}$ by applying an input $u_i$. The resultant states at step $s_{i+1}$ will not necessarily attain the quantized values of states $v$ and $w$. Therefore, we implemented a stochastic transition to the neighboring nodes on the $v-w$ plane as described in \cite{pieterabbeel}. The transition cost is calculated as a weighted sum of the cost at the neighboring nodes, and then stored in the closest node.

Initially, the cyclist starts from the minimum velocity, and his/her remaining energy is initialized at $AWC$. The DP simulation was run on a desktop computer with a 3.2GHz Intel core i5 CPU, and 12GB of RAM. The run time for our longest cycling route (18km) was 53 min and 45 sec. This is thanks to the PMP insights on limited optimal control modes that allowed us to significantly reduce the input quantization. Without it, the computation time would have been in the order of several hours.

\section{Results and Discussion}
\label{sec_optimal_DP_simulation}
\subsection{Simulating Optimal Pacing}
\label{subsec_three_courses}


To illustrate the nature of optimal solution, we first present the optimal pace calculated by DP over 4 km roads with three basic elevation profiles: a flat road, a road with a 5\% grade climb, and a road with two hills. Figure \ref{fig_sim_result_test} shows the three scenarios and the results based on Sub 14 data. The optimal power $u$, remaining anaerobic energy $w$, and velocity $v$ are shown in each case. The figure also shows which one of the four optimal control mode was chosen in the $u_{mode}$ subplot, in which values of 1, 2, 3, and 4 correspond respectively to the control u equaling, 0, $CP$, $u_{\dot{v}=0}$, and $u_{max}$.



       

On the flat road, the optimal strategy is to go all-out till all anaerobic reserves are depleted and then continue with critical power. In this case, the maximum power constraint is activated and the power trajectory stays on the constraint. More interestingly, during the ramp course, the optimal pacing strategy benefits from elevation preview. The cyclist burns only 2\% of anaerobic energy in the starting line to get to the velocity that can be then maintained by applying $CP$. The reserved energy comes to use during the ascent over the hill, by the end of which all of the $AWC$ is expended and the cyclist pedals at $CP$ towards the end. 

During the simulation over a course with two uphill sections, aware of the upcoming downhill, the controller recommends the cyclist to burn most of the anaerobic energy. Then during the downhill, the subject recovers 8\% of the $AWC$ which helps overcoming the second hill. In this case, all four modes of optimal power are applied. We can observe several quick changes in the recommended power level in this scenario. It is important to note that the cyclist is not asked to switch between arbitrary power levels. Following a constant power at $CP$ or a constant velocity during a singular interval is not a difficult task for seasoned cyclists if they are provided with their real-time power and velocity data. For the two other modes, the subject is required to either apply maximum power or stop pedaling, both of which are practical to perform. If necessary, the regularization weight that penalizes the change in the control input can be increased, noting that trial time will increase.

\subsection{Optimal Pace Versus Experimentally Captured Self-Paced Strategy}
\label{subsec_greenville}
In this section we evaluate simulated performance of Subject 14, when optimally paced via DP, against her experimentally measured self-paced performance when riding the cycling course of the 2019 Duathlon National Championship in Greenville, South Carolina on our CompuTrainer in the laboratory setting\footnote{One of the shortcomings of utilizing the CompuTrainer is that it does not model the aerodynamic drag force during a ride. Instead a constant force of $15.5$ N was added to the resistance of the Computrainer as determined during the calibration process. Also the CompuTrainer cannot accelerate the bicycle in downhill rides as would happen on a real ride and therefore the negative resistance force due to downhill section of the course was removed. These changes were also accounted for by adjusting the equations of motion in the DP simulation of the optimal pace when comparing to the Computrainer experimental results.}.

\begin{figure}
\centering
\setlength\fboxsep{0pt}
\setlength\fboxrule{0pt}
\setlength\belowcaptionskip{0pt}
\fbox{\input{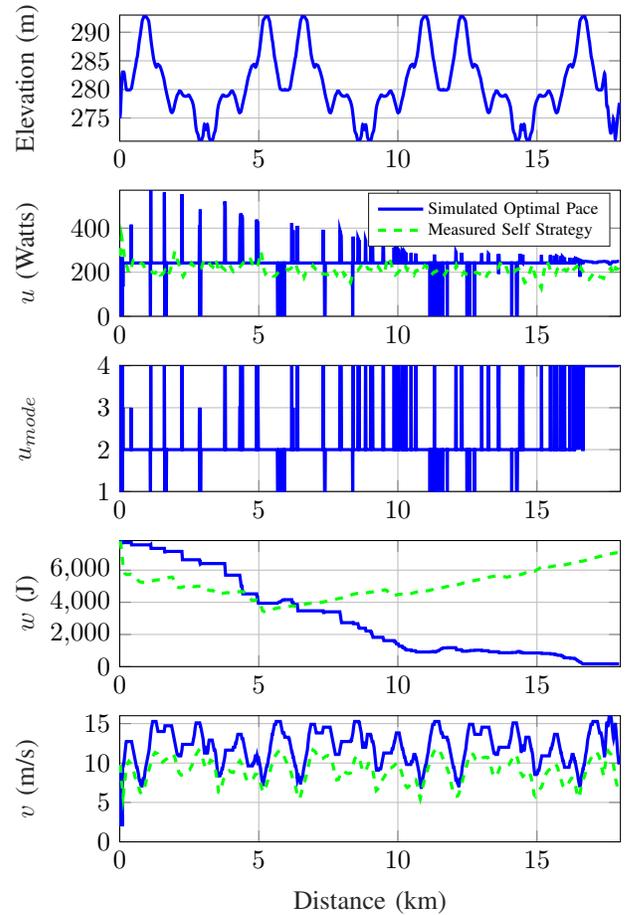}}
\caption{\footnotesize Power and energy trajectory differences between sub 14's self-paced strategy and the optimally-paced simulation on the Greenville Duathlon course.}
\label{fig_Greenville_test}
\end{figure}

We requested Subject 14, who is a seasoned and capable cyclist, to ride the course with her own pacing strategy and asked her to aim to finish in the fastest time possible. During this laboratory experiment, she was able to see in real-time and via the interface depicted in Figure \ref{fig_experiment_setup} the current road slope, distance traveled, and her velocity, pedaling power, and heart rate  The subject managed to finish the course in \textit{34 min 8 sec}. We then evaluated her performance when optimally paced with DP over the same course, in a simulation of her model. Figure \ref{fig_Greenville_test} shows the elevation profile as well as the cyclist pedaling power, remaining anaerobic energy, and velocity in both scenarios. 

The simulation results suggest that if the subject followed the optimal power provided by the DP, she would have finished the course in \textit{25 min 56 sec}, a $24\%$ improvement with respect to her self-paced time-trial. While the large difference can be partially attributed to ideal simulation conditions,  an important observation from this test is that the subject's self strategy lacks consistency in applying power. During the first 5 km of the course in the self-paced scenario, the average power of the cyclist is 230 Watts, whereas during the last 5 km it is 190 Watts. Both of these values are below the Subject's critical power (242 Watts). However, the optimal simulation suggests that except for several short periods of maximal effort and recovery, power should remain at $CP$. While the velocity profile in the simulation is always higher than the subject's velocity, both profiles have a similar trend during the course which is dictated by the elevation profile. The difference between the velocity profiles can be explained by the the short frequent maximal efforts that increases the cyclist's velocity. The average power of the optimal strategy is 240 Watts which is only 28 Watts greater than the average for the self-paced trial.

Also, because the cyclist was pedaling below $CP$ for most parts of the course, her model indicates that she must have recovered almost all of her anaerobic energy at the end which is far from optimal. Although our subject was familiar with the route and cycles four times a week, she is not a professional athlete and lacks high level training, which could be one reason for her lack of pace in spending her anaerobic energy. Nevertheless, pacing over long hilly courses can be a challenge for professional athletes as well, and the proposed optimal pacing strategy could be used for coaching or even real-time guidance of cyclists as will be outlined in Section \ref{pilot_test}. 

\subsection{Sensitivity Analysis}

As mentioned in Section \ref{sec_lit_review}, the 3MT test may overestimate $CP$ and $AWC$. Other studies have shown that $CP$ and $AWC$ can change over a span of four weeks depending on the amount of exercise performed \cite{dekerle2014critical, wright2017reliability}. In this section and via a sensitivity analysis, we evaluate the effect of uncertainty in $CP$ and $AWC$ estimates on the time attained by optimal pacing. We consider i) nominal parameter estimation errors of $\pm 5\%$ as well as ii) erroneous use of parameters of a different subject for subject 14. The best scenario baseline that we compare against is the optimal pace of subject 14 over the 2019 National Duathlon Championship course, as presented in Section \ref{subsec_greenville}. We assume that the baseline simulation uses the true values of parameters $CP$ and $AWC$ when determining the optimal pace. 

In the first set of parameter sensitivity simulations, we evaluate the impact of a $\pm 5\%$ error in estimates of either $CP$ or $AWC$ on the trial time. In other words four scenarios of $\widehat{CP}=(1 \pm 0.05)CP$ and $\widehat{AWC}=(1 \pm 0.05)AWC$ are considered, where $\widehat{CP}$ and $\widehat{AWC}$ denote the estimated values of $CP$ and $AWC$, respectively.  

Since the true values $CP$ and $AWC$ are presumed unknown, the backward DP was run using the estimated values $\widehat{CP}$ and $\widehat{AWC}$ and optimal values of cost-to-go and input were recorded for each $(v,w)$ pair accordingly. In the forward DP calculations, two models were simulated: i) one model with true parameter values to represent the true cyclist, her actual states $v$ and $w$, and her maximum power constraint as a function of $v$ and $w$, and ii) another model with estimated parameters to determine an open-loop estimate of the remaining anaerobic energy $\hat{w}$ since we will not have a sensor to measure the true value of $w$ in a real implementation. Then the actual velocity $v$ which can be measured and the open-loop estimate $\hat{w}$ were used to extract the optimal value of power $u$ from the DP lookup table. Note that the extracted power $u$ may violate the true power constraint calculated earlier due to model parameter mismatch; in which case the maximum power is applied instead in the next forward simulation step. In other words if the recommended power exceeds the cyclist capability, the cyclist is only able to apply her maximal power and this is captured in our simulations.


Travel times recorded for each of the four simulations with perturbed parameters are reported in Table  \ref{tbl:sebsitivity-results} as well as the ideal baseline with true parameters. The travel time in all four cases with parameter value perturbation is larger than the baseline case. This is an expected result because DP provides the global optimal solution, and deviations from DP recommendations due to parameter mismatch, yields larger than optimal travel times. That said, the travel time in the sub-optimal simulations was at most 1\% longer than the optimal baseline. This suggests that the optimal controller still performs well in the presence of reasonable measurement errors or week-to-week changes in $CP$ and $AWC$ values.

\begin{table}[h]
\begin{center}
\captionsetup{width= \columnwidth}
\caption{\vspace{0mm}\footnotesize Simulation results for cases with uncertainty in the estimated $CP$ and $AWC$. The fatigue and recovery models of Sub 14 and the elevation profile of the 2019 National Duathlon Championship was used for these trials.} \label{tbl:sebsitivity-results}
{\begin{tabular}{c|c|c||c}
\hline
	 & $CP$ (Watts) & $AWC$ (J) & time (sec) \\
\hline\hline
Baseline  & 242 & 7841 & 1556 \\
$CP(1 + 0.05)$ &254.1 & 7841 &  1567 \\
$CP(1 - 0.05)$ & 222.9 & 7841 &  1572 \\
$AWC(1 + 0.05)$ & 242& 8233 &  1563 \\
$AWC(1 - 0.05)$ & 242 & 7449 &  1576 \\

\hline
\end{tabular}}

\end{center}
\end{table}

\begin{figure}
\centering
\setlength\fboxsep{0pt}
\setlength\fboxrule{0pt}
\setlength\belowcaptionskip{0pt}
\fbox{\input{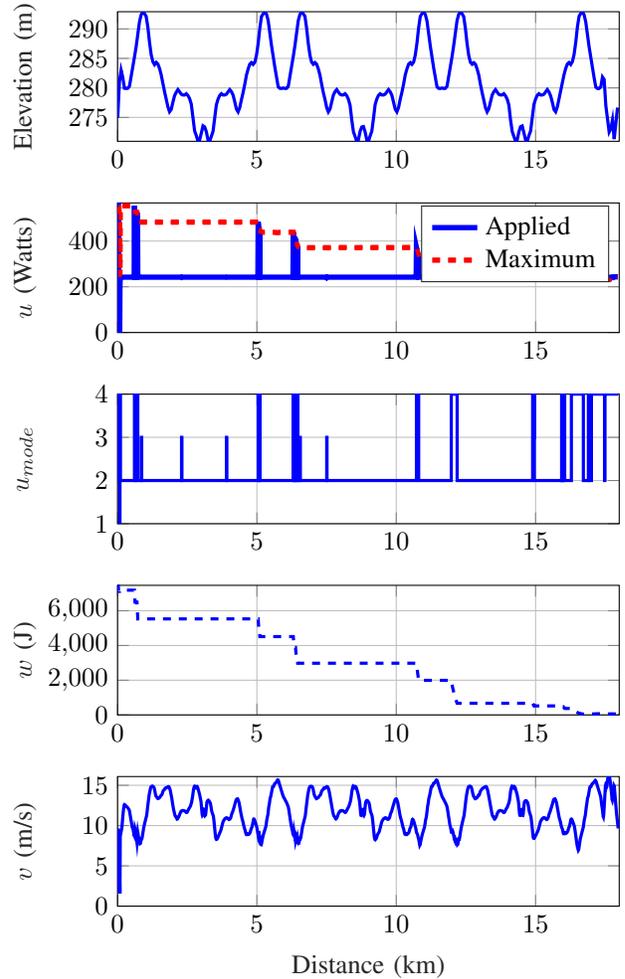}}
\caption{\footnotesize The sensitivity simulation case on the Greenville Duathlon course assuming $AWC$ fro Sub 14 was 5\% overestimated.}
\label{fig_sensitivity}
\end{figure}

Figure \ref{fig_sensitivity} demonstrates the final outcome of a sensitivity simulation in which  $AWC$ had been overestimated by 5\%. Comparing the results shown in this figure to Figure \ref{fig_Greenville_test}, we observe less sprint and no recovery intervals in power trajectory, which results in sub-optimal performance. 
To understand sensitivity of the optimal pace to larger errors in parameter estimate, we evaluated a scenario in which the parameter for a difference subject (subject 6) were erroneously used to determine the pace of subject 14. Similarly to the previous sensitivity study, we used two models for DP calculations. The trial time of subject 14 increased to 1659 seconds which is 7\% higher than the optimal baseline simulation for Sub 14 and could have easily cost her a race. The significant increase in trial time in this case underlines the necessity of using individualized parameters of fatigue and recovery for effective optimal pacing.


\subsection{Laboratory Pilot Test via a Virtual Coach}
\label{pilot_test}

The ultimate goal of this research is to be able to pace a cyclist on a real course by displaying the optimal power to hold at each time . Towards that goal, we have created a laboratory test environment where a cyclist on a stationary CompuTrainer bicycle receives optimal power suggestions on a display in real-time. Here we describe the process for displaying the optimal power to a cyclist and the outcome of a preliminary experiment. 

\begin{figure}
\centering
\setlength\fboxsep{0pt}
\setlength\fboxrule{0pt}
\fbox{\includegraphics[width=\columnwidth]{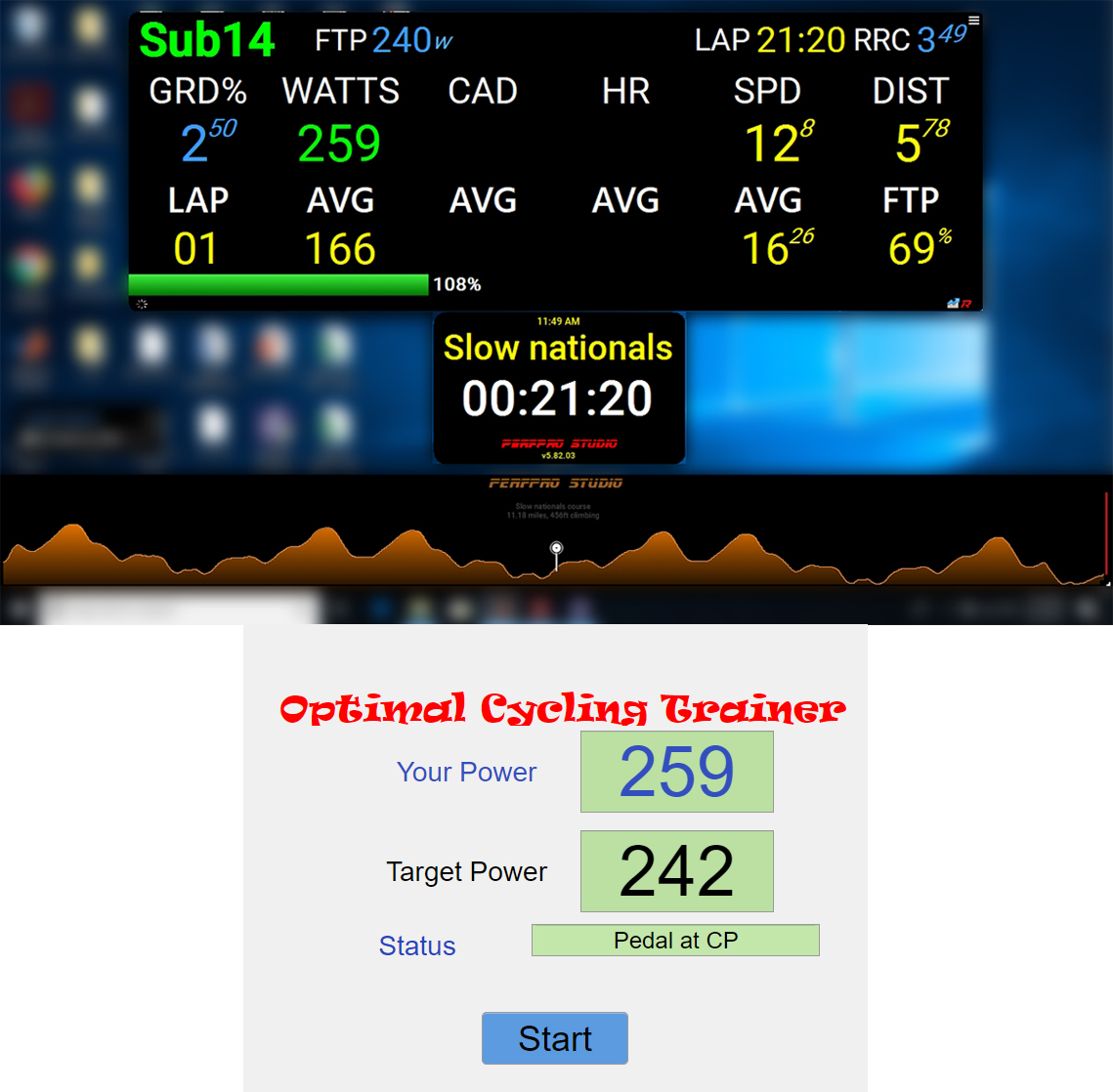}}
\caption{\footnotesize The PerfPro Studio interface which provides real-time data from the CompuTrainer on the top. Our virtual coach interface designed to show the optimal power and real-time power to the cyclist at the bottom.}
\label{fig_software}
\end{figure}

As shown on top of Figure \ref{fig_software}, our CompuTrainer provided a visual display of cyclist's power, velocity, road grade, and position. However our Computrainer did not provide a live feed of this data. Instead we devised an image processing algorithm to  automatically extract the cyclist's power, velocity, and position from the Computrainer screen at 2 Hz. Subsequently, the remaining anaerobic energy $w$ was estimated open-loop and using the fatigue and recovery model of the cyclist. In the future it may be possible to estimate $w$ in closed-loop using non-invasive wearable sensors such as a sweat lactate sensor developed in \cite{nagamine2019noninvasive}.

The road profile was assumed to be available to our \textit{virtual coach} in advance. The backward DP, using parameter models of the cyclist, was run in advance of the cycling session to generate a lookup table of optimal power as a function of the cyclist's states $v$ and $w$ for each position interval of the course.
During the cycling session, the optimal power to hold was retrieved from this table at each position interval, using the estimated state of the cyclist (remaining anaerobic energy and velocity). The suggested power was calculated once every 100 m and was maintained constant during the next position interval, which was comfortable for the cyclist to follow and did not have the rapid changes as those that were observed in previous simulations. The suggested (target) power was displayed in a MATLAB graphic user interface as seen in the bottom of Figure \ref{fig_software}. This interface also displayed the optimal pedaling mode (of the 4 optimal choices described earlier) and the cyclist's actual pedaling power. To ensure that the cyclist only relied on the suggested optimal pace and did not self-strategize, s/he was only allowed to look at the MATLAB interface and not the CompuTrainer's user interface. Therefore, the cyclist did not know the road grade, position, and velocity.

We completed a pilot test using the elevation profile of the 2019 National Duathlon Championship in which subject 14 was instructed to follow the power suggestions displayed to her in real-time. The cyclist was cheered forward by two team members standing alongside as she was also cheered when she had completed the self-paced strategy that was shown in Figure \ref{fig_Greenville_test}. Her trial-time in the pilot trial improved by 3\% over her self-strategy trial presented in Section \ref{subsec_greenville} which is an encouraging result. However we found later and when the subject was not available for further testing, that due to a programming error the suggested pace was sub-optimal and thus cannot make any general conclusion about the result. We expect the performance to further improve with the correction that has been done since. By repeat experiments on several new subjects in the future, we hope to verify that performance can be indeed improved via optimal strategy feedback in the lab and on the road and across different cyclists. We are including this section on the pilot test process in the paper, hoping that the description of the process by itself motivates and helps future work by other groups as well.

\section{Conclusions \& Future Work}

Optimal pacing of a cyclist in a time-trial was formulated as an optimal control problem with an emphasis on the influence of depletion and recovery of anaerobic reserve on performance. In particular each cyclist's maximal power varies with level of fatigue and chosen cadence and plays a critical role in determining the optimal pace in a hilly time-trial. To that end, state-space dynamic models that track depletion and recovery of Anaerobic Work Capacity ($AWC$) as a function of rider's power above or below their Critical Power ($CP$) were hypothesized and were calibrated using data from six human subject tests after each spent 14 hours in the lab. In addition, a model was obtained from experimental data relating cyclists' maximal power to their remaining anaerobic capacity and their cadence. With the models in place, we were able to usefully employ Pontryagin's Minimum Principle to determine that over any road profile, the optimal strategy is bang-singular-bang, switching between maximum exertion, no exertion, pedaling at $CP$, or cruising at constant speeds. Global optimality of these four modes was further verified via the numerical method of Dynamic Programming. Once global optimality was confirmed, limiting the quantization of the cyclist power to only these four optimal modes, substantially reduced the computational load of DP and allowed finer quantization of the states. Simulations over simple road profiles with one or two steep climbs, showed the efficacy of the optimal strategy in distributing the depletion of $AWC$ along the whole course and appropriately pacing the cyclist in anticipation of climbs. We also simulated human subject 14 over the 2019 Greenville Duathlon course. We had data from subject 14 pedaling the same course on a CompuTrainer with her self-strategy. The simulation suggested that optimal pacing could speed up subject 14 and allow her to finish the course in 76\% of her original time (reduction from 2048 seconds to 1556 seconds) with a 28 Watt increase in her average output power which is 11\% of her $CP$. The comparison indicates that consistency in power generation is a key difference between the subject’s self and the optimal strategies. While the power should be kept at $CP$ for most of the ride, the cyclist benefits from frequent short sprints to increase velocity.

We recently completed a pilot test in which subject 14 was instructed to follow a \textit{sub-optimal} strategy displayed to her in real-time and as a result her trial-time improved over self-strategy. We hope to verify that performance can be indeed improved via optimal strategy feedback in repeat experiments once we are able to resume human subject testing in the lab.  Other future work include additional 3MT tests at different fatigue and cadence levels to further validate our maximal power surface depicted in Figure \ref{fig_const_cadence}. A few recent studies \cite{bartram2017predicting, wright2017reliability} suggest that 3MT tests may not provide very accurate estimation of $CP$ and $AWC$ and therefore there is room for additional work on protocols that better estimate these parameters. Our optimal control analysis and strategy do not depend on the value of these parameters and could be exercised with more accurate parameter values.


\section*{ACKNOWLEDGMENT}
The authors would like to thank Furman University students Frank Lara, Mason Coppi, Nicholas Hayden, Lee Shearer, Brendan Rhim, Alec Whyte, and Jake Ogden for their contribution in data collection from recruited human subjects. Also they would like to thank the fantastic subjects who agreed to do a series of extremely challenging experiments and helped this project progress.

\ifCLASSOPTIONcaptionsoff
  \newpage
\fi
\bibliographystyle{IEEEtran}
\bibliography{root}

\vfill\null

\newpage
\begin{IEEEbiography}[{\includegraphics[width=1in,height=1.25in,clip,keepaspectratio]{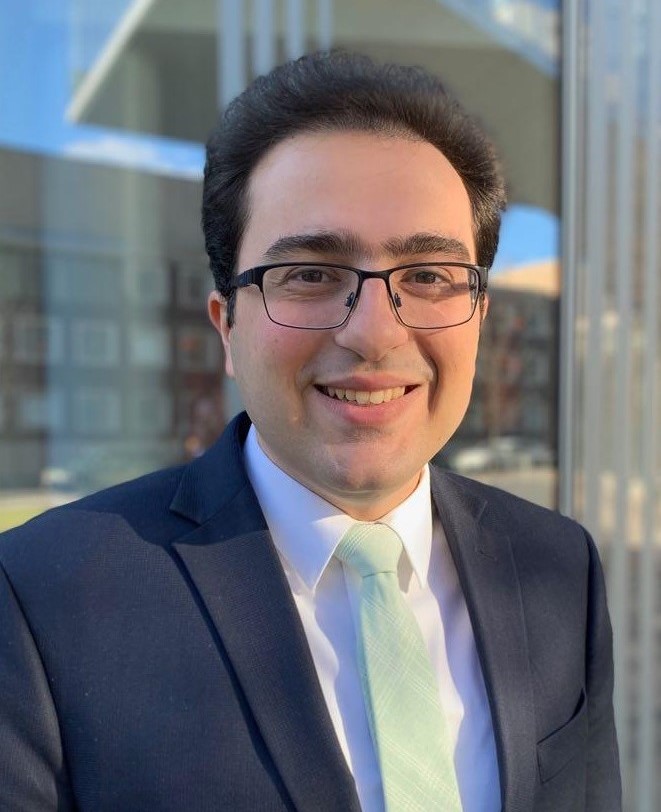}}]{Faraz Ashtiani}
received the Ph.D. degree in mechanical engineering from Clemson University, SC, USA in 2021. He had received his B.Sc. in mechanical engineering from University of Tehran, Iran in 2015. His research interests are the applications of optimal control, intelligent transportation systems, and connected and autonomous vehicle technologies.
\end{IEEEbiography}

\vspace{-10 mm}

\begin{IEEEbiography}[{\includegraphics[width=1in,height=1.25in,clip,keepaspectratio]{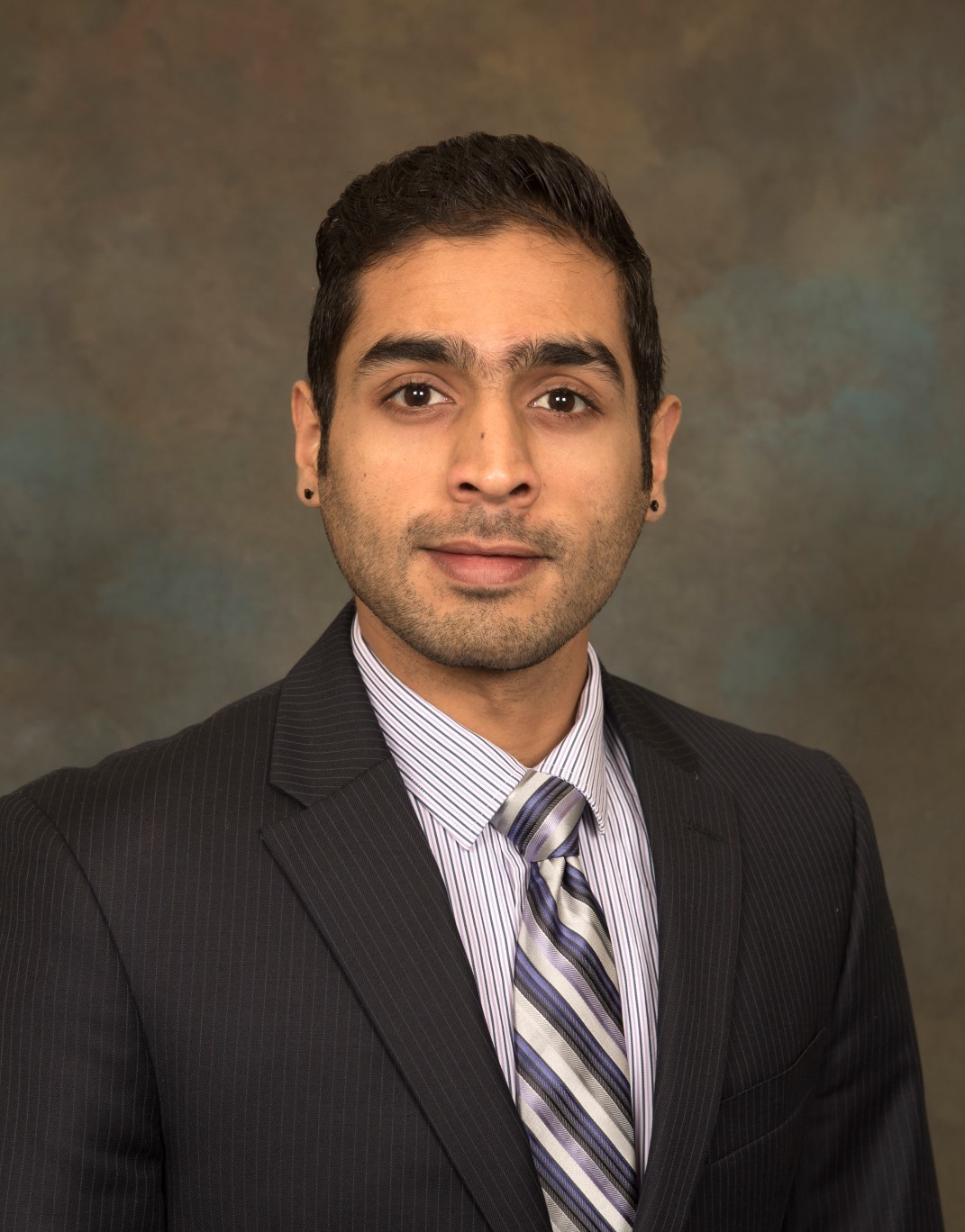}}]{Vijay Sarthy M Sreedhara}
received the Ph.D. degree in mechanical engineering from Clemson University, SC, USA in 2020. He had received his M.Sc. in mechanical engineering from Clemson University, USA in 2015 and his B.E. of Engineering degree in mechanical engineering from Visvesvaraya Technological University, India in 2009. Prior to graduate school, he had worked as an Engineer at John Deere and Toyota in India. His research interests span the areas of human performance optimization, engineering of sport, and product design and development.
\end{IEEEbiography}

\vspace{-10 mm}

\begin{IEEEbiography}[{\includegraphics[width=1in,height=1.25in,clip,keepaspectratio]{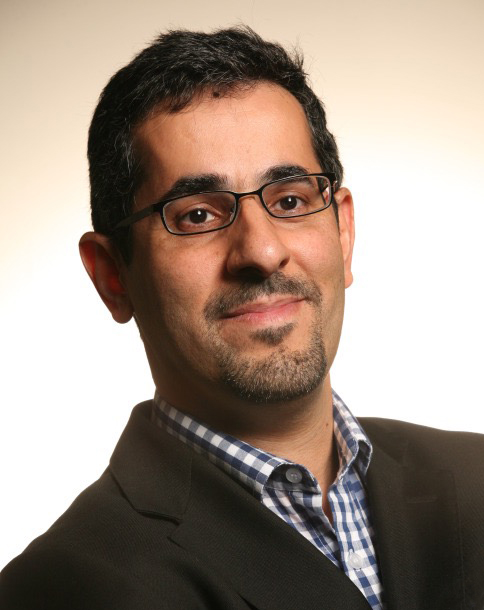}}]{Ardalan Vahidi} 
is a Professor of Mechanical Engineering at Clemson University, South Carolina. He received his Ph.D. in mechanical engineering from the University of Michigan, Ann Arbor, in 2005, M.Sc. in transportation safety from George Washington University, Washington DC, in 2002, and B.S. and M.Sc. in civil engineering from Sharif University, Tehran in 1996 and 1998, respectively. In 2012–2013 he was a Visiting Scholar at the University of California, Berkeley. He has also held scientific visiting positions at BMW Technology Office in California, and at IFP Energies Nouvelles, in France. His research is at the intersection of energy, vehicular systems, and automatic control. His recent publications span topics in optimal motion planning for connected and automated vehicles and intelligent transportation systems.
\end{IEEEbiography}

\vspace{-10 mm}

\begin{IEEEbiography}[{\includegraphics[width=1in,height=1.25in,clip,keepaspectratio]{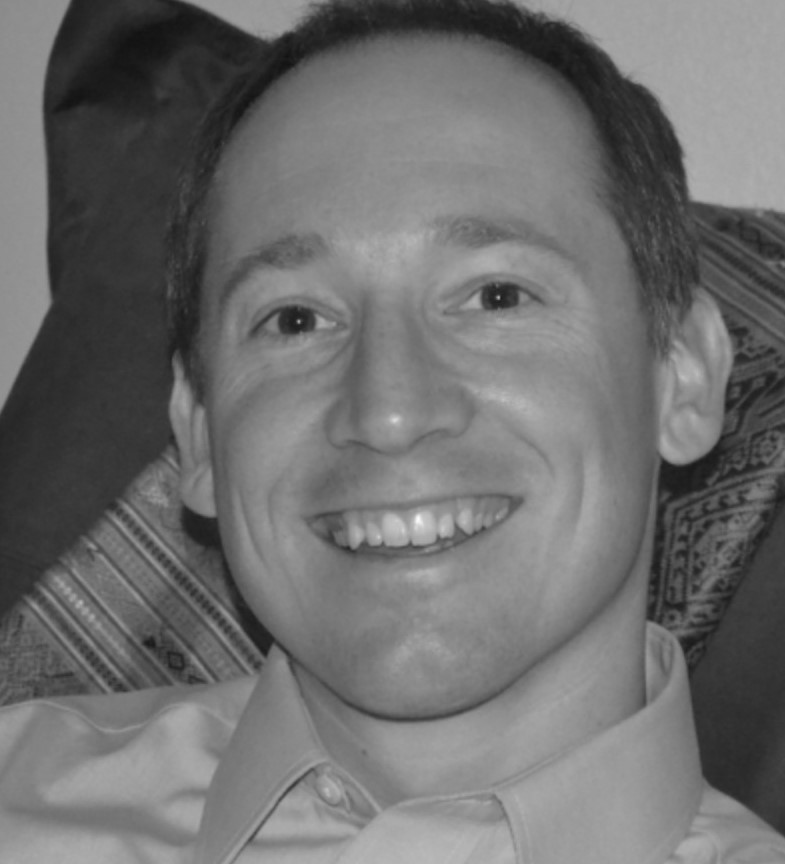}}]{Randolph Hutchison}
received the B.S. degree in aerospace engineering from Virginia Tech University, Blacksburg, VA, USA, in 1999, and both the M.S. and Ph.D. degrees in bioengineering from Clemson University, Clemson, SC, USA, in 2011. He has held positions at Pratt \& Whitney Aircraft Engines as a turbine designer in structural vibrations and aerodynamics in West Palm Beach, FL, and North Haven, CT, USA. He is currently an Associate Professor at Furman University in the Department of Health Sciences, Greenville, SC, USA. His research interests include biomechanical and physiological modeling to optimize human locomotion with the use of wearable sensors. His recent publications span topics in validation of sensors for the measurement of human motion, biomechanics-based orthopaedic rehabilitation, and evaluation of novel power-based training programs.
\end{IEEEbiography}

\vspace{-10 mm}

\begin{IEEEbiography}[{\includegraphics[width=1in,height=1.25in,clip,keepaspectratio]{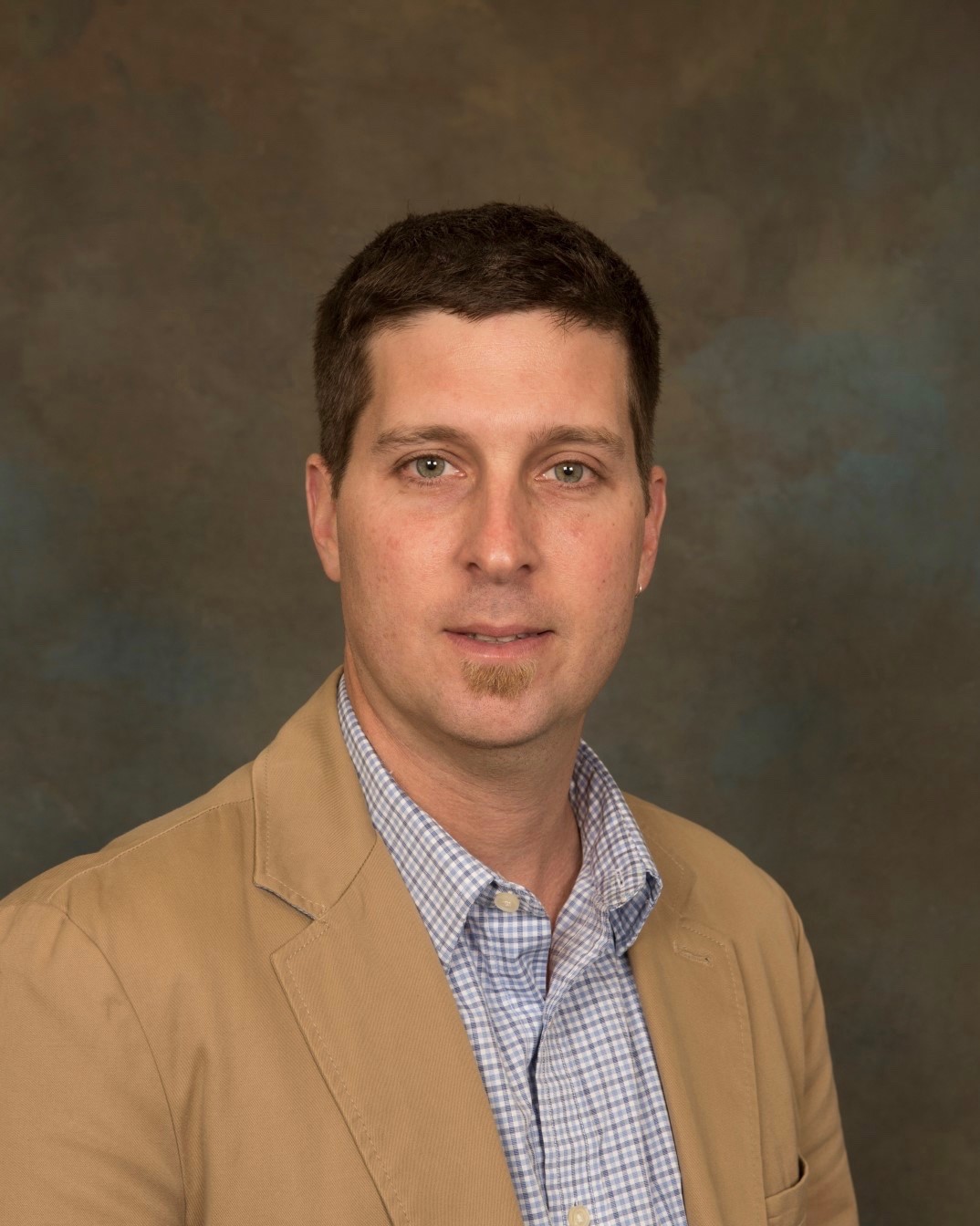}}]{Gregory Mocko}
is currently an Associated Professor in mechanical engineering at Clemson University Clemson SC USA. He received the B.Sc. degree in mechanical engineering and material Science from University of Connecticut, Storrs, CT USA in 1999, M.Sc. degree in mechanical engineering from Oregon State University, Corvallis, OR USA in 2001 and Ph.D. in mechanical engineering from the Georgia Institute of Technology, Atlanta GA USA in 2006. His research interests are broad including human performance, teaming and communication within complex design, and knowledge-based manufacturing. His recent publications include energy expenditure during physical activities, simulation of manufacturing processes, and retrieval of knowledge to support design and manufacturing.
\end{IEEEbiography}


\end{document}